\documentclass[3p]{elsarticle}
\usepackage{makeidx}
\usepackage{amsfonts}
\usepackage{amsmath}
\usepackage{amssymb}
\usepackage{graphicx}
\usepackage{algorithmic}
\usepackage{algorithm}
\usepackage{comment}
\usepackage{hyperref}

\newcommand{\heading}[1]{\medskip\par\noindent{\bf #1}}

\newenvironment{packed_enum}{
	\begin{enumerate}
		\setlength{\itemsep}{1pt}
	    \setlength{\parskip}{0pt}
		\setlength{\parsep}{0pt}
}{\end{enumerate}}

\newenvironment{packed_itemize}{
	\begin{itemize}
		\setlength{\itemsep}{1pt}
	    \setlength{\parskip}{0pt}
		\setlength{\parsep}{0pt}
}{\end{itemize}}

\newtheorem{theorem}{Theorem}[section]
\newtheorem{lemma}[theorem]{Lemma}
\newtheorem{proposition}[theorem]{Proposition}
\newtheorem{corollary}[theorem]{Corollary}

\newtheorem{problem}{Problem}
\newproof{proof}{Proof}

\def\eps{\varepsilon}

\def\O{\mathcal{O}{}}
\def\blt{\blacktriangleleft}

\def\wlt{\vartriangleleft}

\def\lbound{\hbox{\rm{lbound}}}

\def\ubound{\hbox{\rm{ubound}}}
\def\leftshift{\hbox{\rm \sc LeftShift}}
\def\lshift{\text{LS}}
\def\rshift{\text{RS}}
\def\rset{{\mathfrak{Rep}}}

\def\computationproblem#1#2#3#4{
	\vskip 1ex
	\begin{center}
	\fbox{\begin{tabular}{rp{#4}}
	{\bf Problem:\enspace}&#1\\
	{\bf Input:\enspace}&#2\\
	{\bf Output:\enspace}&#3\\
	\end{tabular}}
	\end{center}
	\vskip 1ex
}

\def\eps{\varepsilon}

\def\ext{\textsc{RepExt}}
\def\brep{\textsc{BoundRep}}
\def\simrep{\textsc{SimRep}}
\def\partition{\textsc{3-Partition}}

\def\int{\hbox{\bf \rm \sffamily INT}}
\def\pint{\hbox{\bf \rm \sffamily PROPER INT}}
\def\uint{\hbox{\bf \rm \sffamily UNIT INT}}

\def\ca{\hbox{\bf \rm \sffamily CIRCULAR-ARC}}

\def\cP{\hbox{\rm \sffamily P}}
\def\cNP{\hbox{\rm \sffamily NP}}

  \def\calC{{\cal C}}

 \def\calR{{\cal R}}

\begin{document}

\title{Extending Partial Representations of Proper and Unit Interval Graphs\tnoteref{support}}
\tnotetext[support]{The conference version of this paper appeared in SWAT 2014~\cite{kkorssv}.  The
first, second and sixth authors are supported by ESF Eurogiga project GraDR as GA\v{C}R GIG/11/E023,
the first author also by GA\v{C}R 14-14179S and the first two authors by Charles University as GAUK
196213.  The fourth author is supported by a fellowship within the Postdoc-Program of the German
Academic Exchange Service (DAAD), the sixth author by projects NEXLIZ - CZ.1.07/2.3.00/30.0038,
which is co-financed by the European Social Fund and the state budget of the Czech Republic, and ESF
EuroGIGA project ComPoSe as F.R.S.-FNRS - EUROGIGA NR 13604.}

\author[cunicsi]{Pavel Klav\'{\i}k}
\ead{klavik@iuuk.mff.cuni.cz}
\author[cunidam]{Jan Kratochv\'{\i}l}
\ead{honza@kam.mff.cuni.cz}
\author[jaist]{Yota Otachi}
\ead{otachi@jaist.ac.jp}
\author[cunidam,kit]{Ignaz Rutter}
\ead{rutter@kit.edu}
\author[kobe]{Toshiki Saitoh}
\ead{saitoh@eedept.kobe-u.ac.jp}
\author[uwb]{Maria Saumell}
\ead{saumell@kma.zcu.cz}
\author[cunidam]{Tom\'a\v{s}~Vysko\v{c}il}
\ead{whisky@kam.mff.cuni.cz}

\address[cunicsi]{Computer Science Institute, Faculty of Mathematics and
   		Physics,\\Charles University in Prague, Malostransk{\'e} n{\'a}m{\v e}st{\'\i} 25,
        118 00 Prague, Czech Republic.}
\address[cunidam]{Department of Applied Mathematics, Faculty of Mathematics and
	   	Physics,\\Charles University in Prague, Malostransk{\'e} n{\'a}m{\v e}st{\'\i} 25,
        118 00 Prague, Czech Republic.}
\address[jaist]{School of Information Science, Japan Advanced Institute of
		Science and Technology.\\Asahidai 1-1, Nomi, Ishikawa 923-1292, Japan.}
\address[kobe]{Graduate School of Engineering, Kobe University,\\Rokkodai 1-1, Nada, Kobe, 657-8501,
		Japan.}
\address[uwb]{Department of Mathematics and European Centre of Excellence NTIS (New Technologies for
		the Information Society), University of West Bohemia, Univerzitn\'{\i} 22, 306 14 Plze\v{n},
		Czech Republic.}
\address[kit]{Faculty of Informatics, Karlsruhe Institute of Technology, Fasanengarten 5, 76128 Karlsruhe,
		Germany.}

\begin{abstract}
The recently introduced problem of extending partial interval representations asks, for an interval
graph with some intervals pre-drawn by the input, whether the partial representation can be extended
to a representation of the entire graph.  In this paper, we give a linear-time algorithm for
extending proper interval representations and an almost quadratic-time algorithm for extending unit
interval representations.

We also introduce the more general problem of \emph{bounded representations} of unit interval
graphs, where the input constrains the positions of some intervals by lower and upper bounds.  We
show that this problem is \cNP-complete for disconnected input graphs and give a polynomial-time
algorithm for the special class of instances, where the ordering of the connected components of the
input graph along the real line is prescribed.  This includes the case of partial representation
extension.

The hardness result sharply contrasts the recent polynomial-time algorithm for bounded
representations of proper interval graphs [Balko et al. ISAAC'13].  So unless $\text{\cP} =
\text{\cNP}$, proper and unit interval representations have vastly different structure.  This
explains why partial representation extension problems for these different types of representations
require substantially different techniques.
\end{abstract}

\begin{keyword}
intersection representation\sep
partial representation extension\sep
bounded representations\sep
restricted representation\sep
proper interval graph\sep
unit interval graph\sep
linear programming
\end{keyword}

\maketitle

\section{Introduction}

Geometric intersection graphs, and in particular intersection graphs of objects in the plane, have
gained a lot of interest for their practical motivations, algorithmic applications, and interesting
theoretical properties.  Undoubtedly the oldest and the most studied among them are \emph{interval
graphs} (\int), i.e., intersection graphs of intervals on the real line. They were introduced by
H\'ajos~\cite{hajos_interval_graphs} in the 1950's and the first polynomial-time recognition
algorithm appeared already in the early 1960's \cite{gilmore64}. Several linear-time algorithms are
known, see~\cite{PQ_trees,LBFS_int}.  The popularity of this class of graphs is probably best
documented by the fact that Web of Knowledge registers over 300 papers with the words ``interval
graph'' in the title.  For useful overviews of interval graphs and other intersection-defined
classes, see textbooks~\cite{agt,egr}.

Only recently, the following natural generalization of the recognition problem has been
considered~\cite{kkv}.  The input of the \emph{partial representation extension} problem consists of
a graph and a part of the representation and it asks whether it is possible to extend this partial
representation to a representation of the entire graph.  Klav\'ik et al.~\cite{kkv} give a
quadratic-time algorithm for the class of interval graphs and a cubic-time algorithm for the class
of proper interval graphs. Two different linear-time algorithms are given for interval
graphs~\cite{blas_rutter,kkosv}.  There are also polynomial-time algorithms for function and
permutation graphs~\cite{kkkw} as well as for circle graphs~\cite{cfk}.  Chordal graph
representations as intersection graphs of subtrees of a tree~\cite{kkos} and intersection
representations of planar graphs~\cite{int_planar_hard} are mostly hard to extend.

A related line of research is the complex of simultaneous representation problems,
pioneered by Jampani and
Lubiw~\cite{simultaneous_interval_graphs,jl-srpcc-12}, where one seeks
representations of two (or more) input graphs such that vertices
shared by the input graphs are represented identically in each of the
representations.  Although in some cases the problem of finding
simultaneous representations generalizes the partial representation
extension problem, e.g., for interval graphs~\cite{blas_rutter}, this
connection does not hold for all graph classes.  For example,
extending a partial representation of a chordal graph is
\cNP-complete~\cite{kkos}, whereas the corresponding simultaneous
representation problem is polynomial-time solvable~\cite{jl-srpcc-12}.
While a similar reduction as the one from~\cite{blas_rutter} works for
proper interval graphs, we are not aware of a direct relation between
the corresponding problems for unit interval graphs.

\begin{figure}[b]
\centering
\includegraphics{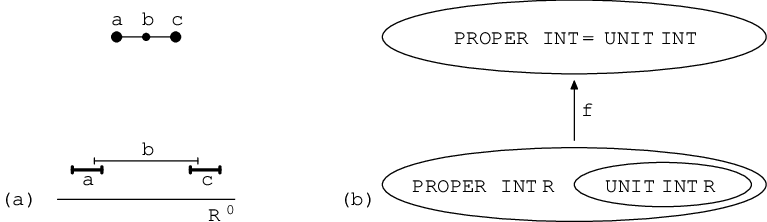}
\caption{(a) A partial representation which is extendible as a proper interval representation, but
not extendible as a unit interval representation. (b) The three structures studied in this paper.
The class of proper/unit interval graphs, all proper interval representations and its substructure of
all unit interval representations. The denoted mapping $f$ assigns to a representation the graph it
represents. The Roberts' Theorem~\cite{proper_is_unit} just states that $f$ restricted to unit
interval representations is surjective.}
\label{fig:pint_and_uint_representations}
\end{figure}

In this paper, we extend the line of research on partial
representation extension problems by studying the corresponding
problems for proper interval graphs (\pint) and unit interval graphs
(\uint).  Roberts' Theorem~\cite{proper_is_unit} states $\pint =
\uint$.  It turns out that specific properties of unit interval
representations were never investigated since it is easier to work
with combinatorially equivalent proper interval representations.  It
is already noted in~\cite{kkv} that partial representation extension
behaves differently for these two classes; see
Figure~\ref{fig:pint_and_uint_representations}a.  This is due to the
fact that for proper interval graphs, in whose representations no interval
is a proper subset of another interval, the extension problem is
essentially topological and can be treated in a purely combinatorial
manner.  On the other hand, unit interval representations, where all
intervals have length one, are inherently geometric, and the
corresponding algorithms have to take geometric constraints into
account.

It has been observed in other contexts that geometric problems are
sometimes more difficult than the corresponding topological problems.
For example, the partial drawing extension of planar graphs is
linear-time solvable~\cite{angelini} for topological drawing but
NP-hard for straight-line drawings~\cite{patrignani}.  Together with
Balko et al.~\cite{bko}, our results show that a generalization of
partial representation extension exhibits this behavior already in
1-dimensional geometry.  The bounded representation problem is
polynomial-time solvable for proper interval graphs~\cite{bko} and
NP-complete for unit interval graphs.  From a perspective of
representations, this result separates proper and unit interval
graphs.  We show that, unless $\text{\cP} = \text{\cNP}$, the
structure of all proper interval representations is significantly
different from the structure of all unit interval representations; see
Figure~\ref{fig:pint_and_uint_representations}b.

Next, we formally introduce the problems we study and describe our
results.

\subsection{Classes and Problems in Consideration}

For a graph $G$, an \emph{intersection representation} $\cal R$ is a collection of sets $\{R_u : u
\in V(G)\}$ such that $R_u \cap R_v \ne \emptyset$ if and only if $uv \in E(G)$; so the edges of $G$
are encoded by the intersections of the sets. An intersection-defined class $\calC$ is the class of all
graphs having intersecting representations with some specific type of sets $R_u$.  For example, in
an \emph{interval representation} each $R_u$ is a closed interval of the real line. A graph is an
\emph{interval graph} if it has an interval representation.

\heading{Studied Classes.}
We consider two classes of graphs. An interval representation is called \emph{proper} if no
interval is a proper subset of another interval (meaning $R_u \subseteq R_v$ implies $R_u = R_v$).
An interval representation is called \emph{unit} if the length of each interval is one. The class of
\emph{proper interval graphs} (\pint) consists of all interval graphs having proper interval
representations, whereas the class of \emph{unit interval graphs} (\uint) consists of all interval
graphs having unit interval representations. Clearly, every unit interval representation is also a
proper interval representation.

In an interval representation $\calR=\{R_v : v\in V\}$, we denote the left and right endpoint of the
interval $R_v$ by $\ell_v$ and $r_v$, respectively.  For numbered vertices $v_1,\dots,v_n$, we
denote these endpoints by~$\ell_i$ and $r_i$.  Note that several intervals may share an endpoint in a
representation. When we work with multiple representations, we use $\calR'$ and $\bar\calR$ for
them. Their intervals are denoted by $R'_v = [\ell'_v,r'_v]$ and $\bar R_v = [\bar\ell_v,\bar r_v]$.

\heading{Studied Problems.}
The \emph{recognition} problem of a class $\calC$ asks whether an input graph belongs to $\calC$;
that is, whether it has a representation by the specific type of sets $R_u$. We study two
generalizations of this problem: The \emph{partial representation extension problem}, introduced
in~\cite{kkv}, and a new problem called the \emph{bounded representation problem}. 

A \emph{partial representation} $\calR'$ of $G$ is a representation of an induced subgraph $G'$ of
$G$. A vertex in $V(G')$ is called \emph{pre-drawn}. A representation $\calR$ \emph{extends}
$\calR'$ if $R_u = R'_u$ for each $u \in V(G')$.

\computationproblem
{$\ext({\calC})$ (Partial Representation Extension of $\calC$)}
{A graph $G$ with a partial representation $\calR'$.}
{Does $G$ have a representation $\calR$ that extends $\calR'$?}
{7.95cm}

\noindent Suppose, that we are given two rational numbers $\lbound(v_i)$ and $\ubound(v_i)$ for each
vertex $v_i$. A representation $\calR$ is called a \emph{bounded representation} if $\lbound(v_i)
\le \ell_i \le \ubound(v_i)$.

\computationproblem
{$\brep$ (Bounded Representation of \uint)}
{A graph $G$ and two rational numbers $\lbound(v_i)$ and~$\ubound(v_i)$ for each $v_i \in V(G)$.}
{Does $G$ have a bounded unit interval representation?}
{7.95cm}

It is easy to see that \brep\ generalizes $\ext(\uint)$ since we can just put $\lbound(v_i) =
\ubound(v_i) = \ell'_i$ for all pre-drawn vertices, and $\lbound(v_i) = -\infty$, $\ubound(v_i) =
\infty$ for the remaining vertices. 

The bounded representation problem can be considered also for interval
graphs and proper interval graphs, where the left and right endpoints
of the intervals can be restricted individually.  A recent paper of
Balko et al.~\cite{bko} proves that this problem is polynomially
solvable for these classes.  Note that for unit intervals, it suffices
to restrict the left endpoint since~$r_i = \ell_i + 1$.  The
complexity for other classes, e.g. circle graphs, circular-arc graphs,
permutation graphs, is open.

\subsection{Contribution and Outline.}

In this paper we present five results.  The first is a simple linear-time algorithm for \ext(\pint),
improving over a previous~$O(nm)$-time algorithm~\cite{kkv}; it is based on known characterizations,
and we present it in Section~\ref{sec:proper_interval_graphs}.

\begin{theorem}
  \label{thm:pint}
  $\ext(\pint)$ can be solved in time $\O(n+m)$.
\end{theorem}

\noindent We note that this algorithm needs some minor and very natural assumption on the encoding
of the input; see Conclusions for details.

Second, in Section~\ref{sec:bounded_representations}, we give a
reduction from \partition\ to show that \brep\ is \cNP-complete for
disconnected graphs.  The main idea is that prescribed intervals
partition the real line into gaps of a fixed width. Integers are
encoded in connected components whose unit interval representations
require a certain width.  By suitably choosing the lower and upper
bounds, we enforce that the connected components have to be placed
inside the gaps such that they do not overlap.

\begin{theorem} \label{thm:brep_npc}
\brep\ is\/ \cNP-complete.
\end{theorem}

Third, in Section~\ref{subsec:lp_approach}, we give a relatively simple quadratic-time algorithm for
the special case of \brep\ where the order of the connected components along the real line is fixed.
We formulate this problem as a sequence of linear programs, and we show that each linear program
reduces to a shortest-path problem which we solve with the Bellmann-Ford algorithm.

The running time is~$O(n^2r + nD(r))$, where~$r$ is the total encoding
length of the bounds in the input, and~$D(r)$ is the time required for
multiplying or dividing two numbers whose binary representation has
length~$r$.  This is due to the fact that the numbers specifying the
upper and lower bounds for the intervals can be quite close to each
other, requiring that the corresponding rationals have an encoding
that is super-polynomial in~$n$.  Clearly, two binary numbers whose
representations have length~$r$ can be added in~$O(r)$ time,
explaining the term of~$O(n^2r)$ in the running time.  However, using
Bellmann-Ford for solving the LP requires also the comparison of
rational numbers.  To be able to do this efficiently, we convert the
rational numbers to a common denominator.  Hence, the multiplication
cost~$D(r)$ enters the running time.  The best known algorithm
achieves $D(r) = \O(r \log r 2^{\log^*r})$~\cite{division}.

Fourth, in
Sections~\ref{subsec:poset_rset}--\ref{subsec:quadratic_algo}, we show
how to reduce the dependency on~$r$ to obtain a running time of~$O(n^2
+ n D(r))$, which may be beneficial for instances with bounds that
have a long encoding.

\begin{theorem} \label{thm:quadratic_brep} \brep\ with a prescribed
  ordering $\blt$ of the connected components can be solved in time
  $\O(n^2+nD(r))$, where $r$ is the size of the input describing bound
  constraints.
\end{theorem}

\noindent Our algorithm is based on shifting intervals.  It starts
with some initial representation and creates, by a series of
transformations, the so-called \emph{left-most representation} of the
input graph.  The algorithm performs $\O(n^2)$ combinatorial
iterations, each taking time $\O(1)$.  The additional time $\O(nD(r))$
is used for arithmetic operations with the bounds.  The main idea for
reducing the running time with respect to the previous approach is to
work with short approximations of the involved rational numbers.  We
compute the precise position of intervals only once, when they reach
their final position.

Further, we derive in Sections~\ref{subsec:eps_grid},
\ref{subsec:poset_rset}, and \ref{subsec:left_shifting} many
structural results concerning unit interval representations. In
particular, we show that all representation of one connected component
form a semilattice.  We believe that these results might be useful in
designing a faster algorithm, attacking other problems, and getting
overall better understanding of unit interval representations.

If the number of connected components is small, we can test all
possible orderings~$\blt$.
\begin{corollary}
  \label{cor:brep_fpt}
  For~$c$ connected components, $\brep$ can be solved in~$O(c! (n^2 +
  nD(r)))$ time.
\end{corollary}

Finally, we note that every instance of \ext(\uint) is an instance of
\brep.  In Section~\ref{sec:unit_interval_graphs}, we show how to
derive for these special instances a suitable ordering~$\blt$ of the
connected components, resulting in an efficient algorithm for
\ext(\uint).

\begin{theorem}
  \label{cor:unit}
  $\ext(\uint)$ can be solved in time $\O(n^2+nD(r))$, where $r$ is
  the size of the input describing positions of pre-drawn intervals.
\end{theorem}

All the algorithms described in this paper are also able to certify the extendibility by
constructing the required representations.

\section{Notation, Preliminaries and Structure} \label{sec:prelim}

As usual, we reserve $n$ for the number of vertices and $m$ for the number of edges of the graph
$G$. We denote the set of vertices by $V(G)$ and the set of edges by $E(G)$. For a vertex $v$, we
denote the closed neighborhood of $v$ by $N[v]=\{x:\, vx\in E(G)\} \cup \{v\}$. We also reserve $r$
for the size of the input describing either bound constraints (for the \brep\ problem) or positions
of pre-drawn intervals (for $\ext(\uint)$). This value $r$ is for the entire graph $G$, and we
use it even when we deal with a single component of $G$. We reserve $c$ for the number of
components of $G$ (maximal connected subgraphs of $G$).

\heading{(Un)located Components.}  Unlike the recognition problem, \ext\ cannot generally be solved
independently for connected components.  A connected component $C$ of~$G$ is \emph{located} if it
contains at least one pre-drawn interval and \emph{unlocated} if it contains no pre-drawn interval.

Let $\calR$ be any interval representation. Then for each component $C$, the union $\bigcup_{u\in C}
R_u$ is a connected segment of the real line, and for different components we get disjoint segments.
These segments are ordered from left to right, which gives a linear ordering $\blt$ of the
components.  So we have $c$ components ordered $C_1 \blt \cdots \blt C_c$.

\heading{Structure.} The main goal of this paper is to establish Theorem~\ref{thm:quadratic_brep}
and to apply it to solve $\ext(\uint)$. Since this paper contains several other results, the
structure might not be completely clear. Now, we try to sketch the story of this paper.

In Section~\ref{sec:proper_interval_graphs}, we describe a key structural lemma of Deng et
al.~\cite{deng}. Using this lemma, we give a simple characterization of extendible instances of
$\ext(\pint)$, which yields the linear-time algorithm of Theorem~\ref{thm:pint}. Also, the reader
gets more familiar with the basic difficulties we need to deal with in the case of unit
interval graphs.

In Section~\ref{sec:bounded_representations}, we show two results for the \brep\ problem. First, we
give a polynomial bound on the required resolution of the drawing. So there exists a value $\eps$,
which is polynomial in the size of the input, such that there exists a representation where, for
every $v_i$, the positions $\ell_i$ and $r_i$ belong to the $\eps$-grid $\{k\eps : k \in \mathbb
Z\}$. Using this, the required representation can be constructed in this $\eps$-grid. Also, we show
that the \brep\ problem is in general \cNP-complete, which proves Theorem~\ref{thm:brep_npc}.

Section~\ref{sec:prescribed_brep} is the main section of this paper and it deals with the \brep\
problem with a prescribed ordering $\blt$ of the components. First, we describe an LP-based
algorithm for solving this problem that solves $2c$ linear programs. Then we derive some structural
results concerning the partially ordered set $\rset$ of all $\eps$-grid unit interval
representations. Using this structure, we conclude the section with a fast combinatorial algorithm
for the above linear programs, solving the \brep\ problem in time $\O(n^2+nD(r))$.

In Section~\ref{sec:unit_interval_graphs}, we show using the main theorem that $\ext(\uint)$ can be
solved in time $\O(n^2+nD(r))$. In Conclusions, we deal with the related problem of simultaneous
representations and give some open problems.


\section{Extending Proper Interval Representations} \label{sec:proper_interval_graphs}

In this section, we describe how to extend partial representations of proper interval graphs in
time $\O(m+n)$. We also give a simple characterization of all extendible instances.

\heading{Indistinguishable Vertices.} Vertices $u$ and $v$ are called \emph{indistinguishable} if
$N[u] = N[v]$. The vertices of $G$ can be partitioned into \emph{groups} of (pairwise)
indistinguishable vertices. Note that indistinguishable vertices may be represented by the same
intervals (and this is actually true for general intersection representations). Since
indistinguishable vertices are not very interesting from the structural point of view, if the
structure of the pre-drawn vertices allows it, we want to \emph{prune} the graph to keep only one
vertex per group.

Suppose that we are given an instance of $\ext(\pint)$. We compute the groups of indistinguishable
vertices in time $\O(n+m)$ using the algorithm of Rose et al.~\cite{recog_chordal_graphs}. Let $u$
and $v$ be two indistinguishable vertices. If $u$ is not pre-drawn, or both vertices are pre-drawn
with $R'_u = R'_v$, then we remove $u$ from the graph, and in the final constructed representation
(if it exists) we put $R_u = R_v$.  For the rest of the section, we shall assume that the input
graph and partial representation are pruned. An important property is that for any representation of
a pruned graph, it holds that all intervals are pairwise distinct.  So if two intervals are
pre-drawn in the same position and the corresponding vertices are not indistinguishable, then we
stop the algorithm because the partial representation is clearly not extendible.

\heading{Left-to-right ordering.}
Roberts~\cite{roberts_phd_thesis} gave the following characterization of proper interval graphs:

\begin{lemma}[Roberts] \label{pint_characterization}
A graph is a proper interval graph if and only if there exists a linear ordering $v_1 \wlt v_2 \wlt
\cdots \wlt v_n$ of its vertices such that the closed neighborhood of every vertex is consecutive.
\end{lemma}

\begin{figure}[t]
\centering
\includegraphics{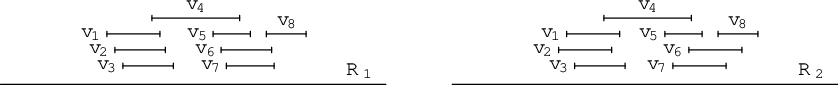}
\caption{Two proper interval representations $\calR_1$ and $\calR_2$ with the left-to-right
orderings $v_1 \wlt v_2 \wlt v_3 \wlt v_4 \wlt v_5 \wlt v_6 \wlt v_7 \wlt v_8$ and $v_2 \wlt v_1
\wlt v_3 \wlt v_4 \wlt v_5 \wlt v_7 \wlt v_6 \wlt v_8$.}
\label{example_ordering}
\end{figure}

This linear order $\wlt$ corresponds to the left-to-right order of the intervals on the real line in
some proper interval representation of the graph. In each representation, the order of the
left endpoints is exactly the same as the order of the right endpoints, and this order satisfies the
condition of Lemma~\ref{pint_characterization}. For an example of $\wlt$, see Figure~\ref{example_ordering}.

How many different orderings $\wlt$ can a proper interval graph admit? In the case of a general
unpruned graph possibly many, but all of them have a very simple structure. In
Figure~\ref{example_ordering}, the graph contains two groups $\{v_1,v_2,v_3\}$ and $\{v_6,v_7\}$.
The vertices of each group have to appear consecutively in the ordering $\wlt$ and may be reordered
arbitrarily. Deng et al.~\cite{deng} proved the following:

\begin{lemma}[Deng et al.] \label{pint_ordering}
For a connected (unpruned) proper interval graph, the ordering $\wlt$ satisfying the condition of
Lemma~\ref{pint_characterization} is uniquely determined up to local reordering
of groups of indistinguishable vertices and complete reversal.
\end{lemma}

\noindent This lemma is key for partial representation extension of proper interval graphs. Essentially, we
just have to deal with a unique ordering (and its reversal) and match the partial representation on
it. Notice that in a pruned graph, if two vertices are indistinguishable, then their order is prescribed by
the partial representation.

We want to construct a partial ordering $<$ which is a simple representation of all orderings $\wlt$
from Lemma~\ref{pint_characterization}. There exists a proper interval representation with an
ordering $\wlt$ if and only if $\wlt$ extends either $<$ or its reversal. According to
Lemma~\ref{pint_ordering}, $<$ can be constructed by taking an arbitrary ordering $\wlt$ and making
indistinguishable vertices incomparable. For the graph in Figure~\ref{example_ordering}, we get
$$(v_1,v_2,v_3) < v_4 < v_5 < (v_6,v_7) < v_8,$$
where groups of indistinguishable vertices are put in brackets. This ordering is unique up to
reversal and can be constructed in time $\O(n+m)$~\cite{uint_corneil}.

\heading{Characterization of Extendible Instances.} We give a simple characterization of the partial
representation instances that are extendible. We start with connected instances. Let $G$ be a pruned
proper interval graph and $\calR'$ be a partial representation of its induced subgraph $G'$. Then
intervals in $\calR'$ are in some left-to-right ordering $<^{\calR'}$. (Recall that the pre-drawn intervals are
pairwise distinct.)

\begin{lemma} \label{lem:pint_conn_extendible}
The partial representation $\calR'$ of a connected graph $G$ is extendible if and only if there
exists a linear ordering $\wlt$ of $V(G)$ such that:
\begin{packed_enum}
\item The ordering $\wlt$ extends $<^{\calR'}$, and either $<$ or its reversal.
\item Let $R'_u$ and $R'_v$ be two pre-drawn touching intervals, i.e., $r_u = \ell_v$, and let $w$
be any vertex distinct from $u$ and $v$. If $uw \in E(G)$, then $w \wlt v$, and if $vw \in E(G)$,
then $u \wlt w$.
\end{packed_enum}
\end{lemma}

\begin{proof}
If there exists a representation $\calR$ extending $\calR'$, then it is in some left-to-right
ordering $\wlt$. Clearly, the pre-drawn intervals are placed the same, so $\wlt$ has to extend
$<^{\calR'}$. According to Lemma~\ref{pint_ordering}, $\wlt$ extends $<$ or its reversal. As for
(2), clearly $v$ has to be the right-most neighbor of $u$ in $\calR$: If $R_w$ is on the right of
$R_v$, it would not intersect $R_u$. Similarly, $u$ is the left-most neighbor of~$v$.

Conversely, let $v_1 \wlt \cdots \wlt v_n$ be an ordering from the statement of the lemma. We
construct a representation $\calR$ extending $\calR'$ as follows. We compute a common linear
ordering $\lessdot$ of the left and right endpoints from left-to-right.\footnote{Notice that, in the
partial representation, some intervals may share position. But if two endpoints $\ell_i$ and $r_j$
share the position, then $v_iv_j \in E(G)$ and we break the tie by setting $\ell_i \lessdot r_j$.}
We start with the ordering $\ell_1 \lessdot \cdots \lessdot \ell_n$, into which we insert the right
endpoints $r_1,\dots,r_n$ one-by-one. For vertex $v_i$, let $v_j$ be its right-most neighbor in the
ordering $\wlt$. Then, we place $r_i$ right before $\ell_{j+1}$ (if $j < n$, otherwise we append
$r_i$ to the end of the ordering).

This left-to-right common order $\lessdot$ is uniquely determined by $\wlt$. Since $\wlt$ extends
$<^{\calR'}$, it is compatible with the partial representation (the pre-drawn endpoints are ordered
as in $\lessdot$). To construct the representation, we just place the non-pre-drawn endpoints
equidistantly into the gaps between neighboring pre-drawn endpoints (or to the left or right of
$\calR'$). It is important that, if two pre-drawn endpoints $\ell_i$ and $r_j$ share their position,
then according to condition (2) there is no endpoint placed in between of $\ell_i$ and $r_j$ in
$\lessdot$ (otherwise one of the two implications would not hold, depending whether a left endpoint
is intersected in between, or a right one). See Figure~\ref{component_representation} for an
example.

\begin{figure}[t]
\centering
\includegraphics{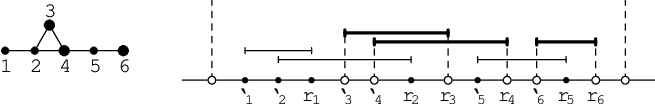}
\caption{Representation of a component with order $1 \wlt 2 \wlt 3 \wlt 4 \wlt 5 \wlt 6$.  First, we
compute the common order of the left and right endpoints:
$\ell_1 \lessdot \ell_2 \lessdot r_1 \lessdot \ell_3 \lessdot \ell_4 \lessdot r_2 \lessdot r_3 \lessdot \ell_5
\lessdot r_4 \lessdot \ell_6 \lessdot r_5 \lessdot r_6$.
The endpoints of the pre-drawn intervals split the segment into several subsegments. We place the
remaining endpoints in this order and, within every subsegment, distributed equidistantly.}
\label{component_representation}
\end{figure}

We argue correctness of the constructed representation $\calR$. First, it extends $\calR'$, since
the pre-drawn intervals are not modified. Second, it is a correct interval representation: Let $v_i$
and $v_j$ be two vertices with $v_i \wlt v_j$, and let $v_k$ be the right-most neighbor of $v_i$ in
$\wlt$. If $v_iv_j \in E(G)$, then $\ell_i \lessdot \ell_k \lessdot r_i$ and, by consecutivity of
$N[u]$ in $\wlt$, we have $\ell_j \lessdot \ell_k$. Therefore, $R_{v_i}$ and $R_{v_j}$ intersect. If
$v_iv_j \notin E(G)$ and $v_j \neq v_{k+1}$, then $r_i \lessdot \ell_{k+1} \lessdot \ell_j$, so
$R_{v_i}$ and $R_{v_j}$ do not intersect. If $v_iv_j \notin E(G)$ and $v_j = v_{k+1}$, then $r_i
\lessdot \ell_{k+1}$ and $R_{v_i}$ and $R_{v_j}$ do not intersect. Finally, we argue that $\calR$ is
a proper interval representation. In $\lessdot$ the order of the left endpoints is the same as the
order of the right-endpoints, since $r_{i+1}$ is always placed on the right of $r_i$ in $\lessdot$.

We conclude that the representation $\calR$ can be made small enough to fit into any open segment of
the real line that contains all pre-drawn intervals.\qed
\end{proof}

Now, we are ready to characterize general solvable instances.

\begin{lemma} \label{lem:pint_extendible}
A partial representation $\calR'$ of a graph $G$ is extendible if and only if
\begin{packed_enum}
\item for each component $C$, the partial representation $\calR'_C$ consisting of the pre-drawn intervals
in $C$ is extendible, and
\item pre-drawn vertices of each component are consecutive in $<^{\calR'}$.
\end{packed_enum}
\end{lemma}

\begin{figure}[t]
\centering
\includegraphics{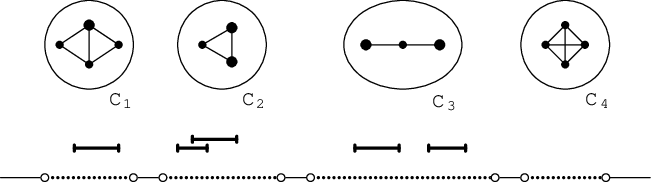}
\caption{An example of a graph with four components $C_1,\dots,C_4$. The pre-drawn intervals give the order
of the located components $C_1 \blt C_2 \blt C_3$. The non-located component $C_4$ is placed to the
right. For each component, we reserve some segment in which we construct the representation.}
\label{component_area_example}
\end{figure}

\begin{proof}
The necessity of (1) is clear. For (2), if some component $C$ would not have its pre-drawn vertices
consecutive in $<^{\calR'}$, then $\bigcup_{u \in C} R_u$ would not be a connected segment of the
real line (contradicting existence of $\blt$ from Preliminaries).

Now, if the instance satisfies both conditions we can construct a correct representation $\calR$
extending $\calR'$ as follows. Using (2), the located components are
ordered from left to right, and we assign pairwise disjoint \emph{open segments} containing all their
pre-drawn intervals (there is a non-empty gap between located components we can use). To unlocated
components, we assign pairwise disjoint open segments to the right of the right-most located
component. See Figure~\ref{component_area_example}. For each component, we construct a
representation in its open segment, using the construction in the proof of
Lemma~\ref{lem:pint_conn_extendible}.\qed
\end{proof}

We are ready to prove that $\ext(\pint)$ can be solved in time $\O(n+m)$:

\begin{proof}[Theorem~\ref{thm:pint}]
We just use the characterization by
Lemma~\ref{lem:pint_extendible}, of which the conditions (1) and
(2) can be easily checked in time $\O(n+m)$. For
Lemma~\ref{lem:pint_conn_extendible}, we check for each component
both constraints (1) and (2). To check (2), we compute for $<$
and its reversal the unique orderings $\wlt$. We test for each of
them whether each touching pair of pre-drawn intervals is placed
in $\wlt$ according to~(2).

If necessary, a representation $\calR$ can be constructed in the same running time since the proofs
of Lemmas~\ref{lem:pint_conn_extendible} and~\ref{lem:pint_extendible} are constructive.\qed
\end{proof}

\section{Bounded Representations of Unit Interval Graphs} \label{sec:bounded_representations}

In this section, we deal with bounded representations. An input of \brep\ consists of a graph $G$
and, for each vertex $v_i$, a \emph{lower bound} $\lbound(v_i)$ and an \emph{upper bound} $\ubound(v_i)$. (We
allow $\lbound(v_i) = -\infty$ and $\ubound(v_i) = +\infty$.) The problem asks whether there exists
a unit interval representation $\calR$ of $G$ such that $\lbound(v_i) \le \ell_i \le \ubound(v_i)$
for each interval $v_i$. Such a representation is called a \emph{bounded representation}.

Since unit interval representations are proper interval representations, all properties of proper
interval representations described in Section~\ref{sec:proper_interval_graphs} hold, in
particular the properties of orderings $\wlt$ and $<$.

\subsection{Representations in $\eps$-grids} \label{subsec:eps_grid}

Endpoints of intervals can be positioned at arbitrary real numbers. For the purpose of the
algorithm, we want to work with representations drawn in limited resolution. For a given instance of
the bounded representation problem, we want to find a lower bound for the required resolution such
that this instance is solvable if and only if it is solvable in this limited resolution.

More precisely, we want to represent all intervals so that their endpoints correspond to points on
some grid. For a value $\eps = {1 \over K} > 0$, where $K$ is an integer, the $\eps$-grid is the set
of points $\{k\eps : k \in \mathbb Z\}$.\footnote{If $\eps$ was not of the form $1 \over K$, then
the grid could not contain both left and right endpoints of the intervals. We reserve $K$ for the
value $1 \over \eps$ in this paper.} For a given instance of \brep, we ask which value of $\eps$
ensures that we can construct a representation having all endpoints on the $\eps$-grid.  So the
value of $\eps$ is the resolution of the drawing.

If there are no bounds, every unit interval graph has a representation in the grid of size $1 \over
n$~\cite{uint_corneil}. In the case of \brep, the size of the grid has to depend on the values of
the bounds. Consider all values $\lbound(v_i)$ and $\ubound(v_i)$ distinct from $\pm\infty$, and
express them as irreducible fractions ${p_1 \over q_1}, {p_2 \over q_2}, \cdots, {p_b \over q_b}$.
Then we define:
\begin{equation} \label{eps_def}
\eps' := {1 \over \text{lcm}(q_1,q_2,\dots,q_b)},\qquad{\text{and}}\qquad\eps := {\eps' \over n},
\end{equation}
where $\text{lcm}(q_1,q_2,\dots,q_b)$ denotes the \emph{least common multiple} of $q_1,\dots,q_b$.
It is important that the size of this $\eps$ written in binary is $\O(r)$. We show that the $\eps$-grid
is sufficient to construct a bounded representation:

\begin{lemma} \label{eps_grid}
If there exists a bounded representation $\calR'$ for an input of the problem \brep, there exists a
bounded representation $\calR$ in which all intervals have endpoints on the $\eps$-grid, where $\eps$
is defined by (\ref{eps_def}).
\end{lemma}

\begin{proof}
We construct an $\eps$-grid representation $\calR$ from $\calR'$ in two steps. First, we shift
intervals to the left, and then we shift intervals slightly back to the right. For every interval
$v_i$, the sizes of the left and right shifts are denoted by $\lshift(v_i)$ and $\rshift(v_i)$
respectively. The shifting process is shown in Figure~\ref{eps_grid_figure}.

\begin{figure}[t!]
\centering
\includegraphics{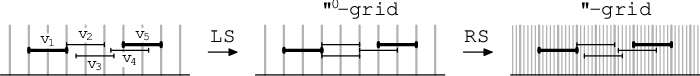}
\caption{In the first step, we shift intervals to the left to the $\eps'$-grid.
The left shifts of $v_1,\dots,v_5$ are $(0,0,{1 \over 2}\eps',{1 \over 3}\eps',0)$.
In the second step, we shift to the right in the refined $\eps$-grid. Right shifts have the same
relative order as left shifts: $(0,0,2\eps,\eps,0)$.}
\label{eps_grid_figure}
\end{figure}

In the first step, we consider the $\eps'$-grid and shift all the intervals to the left to the
closest grid-point (we do not shift an interval if its endpoints are already on the grid).
Original intersections are kept by this shifting, since if $x$ and $y$ are two endpoints satisfying $x
\le y$ before the left-shift, then $x \le y$ also holds after the left-shift. So if $v_iv_j \in E$ and $\ell_i\leq
\ell_j\leq r_i$ before the shift, then these inequalities are preserved by the shifting. On the other hand, we may
introduce additional intersections by shifting two non-intersecting intervals to each other. In this
case, after the left-shift, the intervals only touch; for an example, see vertices $v_2$ and $v_4$ in
Figure~\ref{eps_grid_figure}.

The second step shifts the intervals to the right in the refined $\eps$-grid to remove the additional
intersections created by the first step. The right-shift is a mapping
$$\rshift: \{v_1,\dots,v_n\} \to \{0, \eps, 2\eps, \dots, (n-1)\eps\}$$
having the \emph{right-shift property}: For all pairs $(v_i, v_j)$ with $r_i = \ell_j$,
$\rshift(v_i) \ge \rshift(v_j)$ if and only if $v_iv_j \in E$.  So the right-shift property ensures
that $\rshift$ fixes wrongly represented touching pairs created by $\lshift$.

To construct such a mapping $\rshift$, notice that if we relax the image of $\rshift$ to
$[0,\eps')$, the reversal of $\lshift$ would have the right-shift property, since it produces the
original correct representation $\calR'$. But the right-shift property depends only on the relative
order of the shifts and not on the precise values. Therefore, we can construct $\rshift$ from the
reversal of $\lshift$ by keeping the shifts in the same relative order. If $\lshift(v_i)$ is one of
the $k$th smallest shifts, we set $\rshift(v_i) = (k-1)\eps$.\footnote{In other words, for the
smallest shifts we assign the right-shift $0$; for the second smallest shifts, we assign $\eps$; for
the third smallest shifts, $2\eps$; and so on.} See Figure~\ref{eps_grid_figure}.

We finally argue that these shifts produce a correct $\eps$-grid representation. The right-shift
does not create additional intersections: After $\lshift$ non-intersecting pairs are at distance
at least $\eps'=n\eps$, and by $\rshift$ they can get closer by at most $(n-1)\eps$.
Also, if after $\lshift$ two intervals overlap by at least $\eps'$, their intersection is not
removed by $\rshift$. The only intersections which are modified by $\rshift$ are touching pairs of
intervals $(v_i,v_j)$ having $r_i = \ell_j$ after $\lshift$. The mapping $\rshift$ shifts these
pairs correctly according to the edges of the graph.

Next we look at the bound constraints. If, before the shifting, $v_i$ was satisfying $\ell_i \ge
\lbound(v_i)$, then this is also satisfied after $\lshift(v_i)$ since the $\eps'$-grid contains the
value $\lbound(v_i)$. Obviously, the inequality is not broken after $\rshift(v_i)$. As for the upper
bound, if $\lshift(v_i)=0$ and $\rshift(v_i)=0$, then the bound is trivially satisfied. Otherwise,
after $\lshift(v_i)$ we have $\ell_i \le \ubound(v_i)-\eps'$, so the upper bound still holds after
$\rshift(v_i)$.\qed
\end{proof}

Additionally, Lemma~\ref{eps_grid} shows that it is always possible to construct an $\eps$-grid
representation having the same topology as the original representation, in the sense that overlapping pairs
of intervals keep overlapping, and touching pairs of intervals keep touching.  Also notice that both
representations $\calR$ and $\calR'$ have the same order of the intervals.

In the standard unit interval graph representation problem, no bounds on the positions
of the intervals are given, and we get $\eps' = 1$ and $\eps = {1 \over n}$. Lemma~\ref{eps_grid}
proves in a particularly clean way that the grid of size $1 \over n$ is sufficient to construct
unrestricted representations of unit interval graphs. Corneil et al.~\cite{uint_corneil} show how to
construct this representation directly from the ordering $<$, whereas we use some given
representation to construct an $\eps$-grid representation.

\subsection{Hardness of \brep}

In this subsection we focus on hardness of bounded representations of unit interval
graphs. We prove Theorem~\ref{thm:brep_npc} stating that \brep\ is \cNP-complete.

We reduce the problem from \partition.  An input of \partition\ consists of natural numbers $k$,
$M$, and $A_1,\dots,A_{3k}$ such that ${M \over 4} < A_i < {M \over 2}$ for all $i$, and $\sum A_i =
kM$. The question is whether it is possible to partition the numbers $A_i$ into $k$ triples such
that each triple sums to exactly $M$. This problem is known to be strongly \cNP-complete (even if
all numbers have polynomial sizes)~\cite{partition}.

\begin{proof}[Theorem~\ref{thm:brep_npc}]
According to Lemma~\ref{eps_grid}, if there exists a representation satisfying the bound
constraints, then there also exists an $\eps$-grid representation with this property. Since the
length of $\eps$ given by~(\ref{eps_def}), written in binary, is polynomial in the size of the
input, all endpoints can be placed in polynomially-long positions. Thus we can guess the bounded
representation and the problem belongs to \cNP.

Let us next prove that the problem is \cNP-hard. For a given input of \partition, we construct the
following unit interval graph $G$.  For each number $A_i$, we add a path $P_{2A_i}$ (of length
$2A_i-1$) into $G$ as a separate component.  For all vertices $x$ in these paths, we set bounds
$$\lbound(x) = 1\qquad\text{and}\qquad\ubound(x) = k \cdot (M+2).$$
In addition, we add $k+1$ independent vertices $v_0,v_1,\dots,v_k$, and make their positions in the
representation fixed:
$$\lbound(v_i) = \ubound(v_i) = i \cdot (M + 2).$$
See Figure~\ref{fig:partition_reduction} for an illustration of the reduction. Clearly, the
reduction is polynomial.

\begin{figure}[b]
\centering
\includegraphics{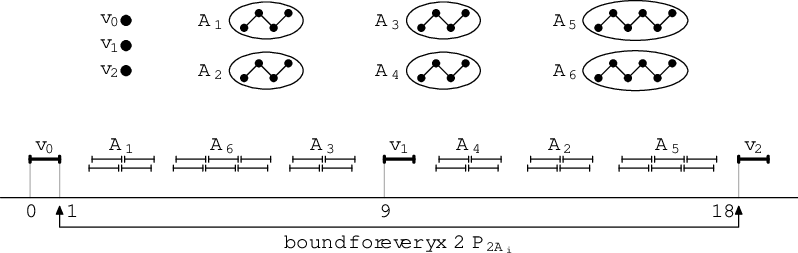}
\caption{We consider the following input for \partition: $k = 2$, $M = 7$, $A_1 = A_2 = A_3 = A_4 =
2$ and $A_5 = A_6 = 3$. The associated unit interval graph is depicted on top, and at the bottom we find
one of its correct bounded representations, giving 3-partitioning $\{A_1,A_3,A_6\}$ and $\{A_2,A_4,A_5\}$.}
\label{fig:partition_reduction}
\end{figure}

We now argue that the bounded representation problem is solvable if and only if the given input of
\partition\ is solvable. Suppose first that the bounded representation problem admits a solution.
There are $k$ gaps between the fixed intervals $v_0,\dots,v_k$ each of which has space less than
$M+1$. (The length of the gap is $M+1$ but the endpoints are taken by $v_i$ and $v_{i+1}$.) The
bounds of the paths force their representations to be inside these gaps, and each path lives in
exactly one gap. Hence the representation induces a partition of the paths.

Now, the path $P_{2A_i}$ needs space at least $A_i$ in every representation since it has an
independent set of the size $A_i$. The representations of the paths may not overlap and the space in
each gap is less than $M+1$, hence the sum of all $A_i$'s in each part is at most $M$. Since the
total sum of $A_i$'s is exactly $kM$, the sum in each part has to be $M$. Thus the obtained
partition solves the \partition\ problem.

Conversely, every solution of \partition\ can be realized in this way.\qed
\end{proof}


\section{Bounded Representations of Unit Interval Graphs with Prescribed Ordering}
\label{sec:prescribed_brep}

In this section, we deal with the \brep\ problem when a fixed ordering $\blt$ of the components is
prescribed.  First we solve the problem using linear programming. Then we describe additional
structure of bounded representations, and using this structure we construct an almost quadratic-time
algorithm that solves the linear programs.

\subsection{LP Approach for \brep} \label{subsec:lp_approach}

\begin{figure}[b]
\centering
\includegraphics{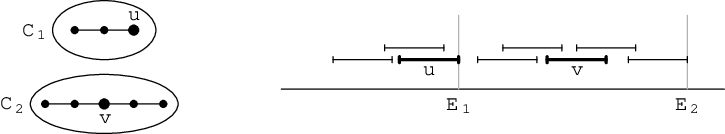}
\caption{The positions of the vertices $u$ and $v$ are fixed by the bound constraints.  The
component $C_1$ can only be represented with $u$ being the right-most interval, since otherwise
$C_1$ would block space for the component $C_2$.}
\label{restricting_components}
\end{figure}

According to Lemma~\ref{pint_ordering}, each component of $G$ can be represented in at most two
different ways, up to local reordering of groups of indistinguishable vertices.  Unlike the case of
proper interval graphs, we cannot arbitrarily choose one of the orderings, since neighboring
components restrict each other's space. For example, only one of the two orderings for the component
$C_1$ in Figure~\ref{restricting_components} makes a representation of~$C_2$ possible.

In the algorithm, we process components $C_1 \blt C_2 \blt \cdots \blt C_c$ from left to right and
construct representations for them. When we process a component $C_t$, we want to represent it on
the right of the previous component $C_{t-1}$, and we want to push the representation of $C_t$ as
far to the left as possible, leaving as much space for $C_{t+1},\dots,C_c$ as possible.

Now, we describe in details, how we process a component $C_t$. We calculate by the algorithm of Corneil et al.
the partial ordering $<$ described in Section~\ref{sec:proper_interval_graphs} and its reversal.
The elements that are incomparable by these partial orderings are vertices of the same group of
indistinguishable vertices. For these vertices, the following holds:

\begin{lemma} \label{lem:group_ordering}
Suppose there exists some bounded representation $\calR$. Then there exists a bounded representation $\calR'$
such that, for every indistinguishable pair $v_i$ and $v_j$ satisfying $\lbound(v_i) \le \lbound(v_j)$, it
holds that $\ell'_i \le \ell'_j$.
\end{lemma}

\begin{proof}
Given a representation $\calR$, we call a pair $(v_i,v_j)$ \emph{bad} if $v_i$ and $v_j$ are
indistinguishable, $\lbound(v_i) \le \lbound(v_j)$ and $\ell_i > \ell_j$.
We describe a process which iteratively constructs $\calR'$ from $\calR$, by constructing a sequence
of representations $\calR = \calR_0,\calR_1,\dots,\calR_k = \calR'$, where the positions in
a representation $\calR_s$ are denoted by $\ell^s_i$'s.

In each step $s$, we create $\calR_s$ from $\calR_{s-1}$ by fixing one bad pair $(v_i,v_j)$: we set
$\ell^{s}_i = \ell^{s-1}_j$ and the rest of the representation remains the same.  Since $v_i$ and $v_j$
are indistinguishable and $\calR_{s-1}$ is correct, the obtained $\calR_s$ is a representation.
Regarding bound constraints,
$$\lbound(v_i) \le \lbound(v_j) \le \ell^{s-1}_j = \ell^s_i < \ell^{s-1}_i \le \ubound(v_i),$$
so the bounds of $v_i$ are satisfied.

Now, in each $\calR_s$ the set of all left endpoints is a subset of the set of all left endpoints of
$\calR$. In each step, we move one left-endpoint to the left, so each endpoint is moved at most
$n-1$ times. Hence the process terminates after $\O(n^2)$ iterations and produces a representation
$\calR'$ without bad pairs as requested.\qed
\end{proof}

For $<$ and its reversal, we use Lemma~\ref{lem:group_ordering} to construct linear orderings
$\wlt$: If $v_i$ and $v_j$ belong to the same group of indistinguishable vertices and $\lbound(v_i)<\lbound(v_j)$,
then $v_i \wlt v_j$. If $\lbound(v_i)=\lbound(v_j)$, we choose any order $\wlt$ between $v_i$ and $v_j$.

We obtain two total orderings $\wlt$, and we solve a linear program for each of them.  Let $v_1 \wlt
v_2 \wlt \cdots \wlt v_k$ be one of these orderings.  We denote the right-most endpoint of a
representation of a component $C_t$ by $E_t$. Additionally, we define $E_0 = -\infty$.
Let $\eps$ be defined as in~(\ref{eps_def}). We modify all lower bounds by putting $\lbound(v_i) =
\max\bigl\{\lbound(v_i),E_{t-1}+\eps\bigr\}$ for every interval $v_i$, which forces the
representation of $C_t$ to be on the right of the previously constructed representation of
$C_{t-1}$. The linear program has variables $\ell_1,\dots,\ell_k$, and it minimizes the value of
$E_t$. We solve:
\begin{alignat}{3}
&\text{Minimize:}\qquad&E_t :\!&= \ell_k+1,\qquad\ \,&&\nonumber\\
&\text{subject to:}\qquad&\ell_i &\le \ell_{i+1},&\qquad&\forall i = 1,\dots,k-1,           \label{constr_ordering}\\
&&\ell_i &\ge \lbound(v_i),&&\forall i = 1,\dots,k,     \label{constr_lb}\\
&&\ell_i &\le \ubound(v_i),&&\forall i = 1,\dots,k,     \label{constr_ub}\\
&&\ell_i &\ge \ell_j - 1,&&\forall v_iv_j \in E(G), v_i \wlt v_j,          \label{constr_edges}\\
&&\ell_i + \eps &\le \ell_j -1,&&\forall v_iv_j \notin E(G), v_i \wlt v_j. \label{constr_nonedges}
\end{alignat}

We solve the same linear program for the other ordering of the vertices of $C_t$. If none of the two
programs is feasible, we report that no bounded representation exists. If exactly one of them is
feasible, we keep the values obtained for $\ell_1,\dots,\ell_k$ and $E_t$, and process the next
component $C_{t+1}$. If the two problems are feasible, we keep the solution in which the value of $E_t$
is smaller, and process $C_{t+1}$.

\begin{lemma} \label{lem:constraints}
Let the representation of $C_{t-1}$ be fixed.  Every bounded $\eps$-grid representation of the
component $C_t$ with the left-to-right order $v_1 \wlt \cdots \wlt v_k$ which is on the right of
the representation of $C_{t-1}$ satisfies constraints (\ref{constr_lb})--(\ref{constr_nonedges}).
\end{lemma}

\begin{proof}
Constraints of types~(\ref{constr_lb}) and~(\ref{constr_ub}) are satisfied, since the representation
is bounded and on the right of $C_{t-1}$. Constraints of type~(\ref{constr_edges}) correspond to
a correct representation of intersecting pairs of intervals. The non-intersecting pairs of an
$\eps$-grid representation are at distance at least $\eps$, which makes constraints of
type~(\ref{constr_nonedges}) satisfied.\qed
\end{proof}

Now, we are ready to show:

\begin{proposition} \label{LP_approach}
The \brep\ problem with prescribed $\blt$ can be solved in polynomial time.
\end{proposition}

\begin{proof}
Concerning the running time, it depends polynomially on the sizes of $n$ and $\eps$, which are
polynomial in the size of the input $r$.  It remains to show correctness.

Suppose that the algorithm returns a candidate for a bounded representation. The formulation of the
linear program ensures that it is a correct representation: Constraints of
type~(\ref{constr_ordering}) make the representation respect $\wlt$. Constraints of
type~(\ref{constr_lb}) and~(\ref{constr_ub}) enforce that the given lower and upper bounds for the
positions of the intervals are satisfied, force the prescribed ordering $\blt$ on the representation
of $G$, and force the drawings of the distinct components to be disjoint. Finally, constraints of
type~(\ref{constr_ordering}), (\ref{constr_edges}) and~(\ref{constr_nonedges}) make the drawing of
the vertices of a particular component $C_t$ to be a correct representation.

Suppose next that a bounded representation exists. According to Lemma~\ref{eps_grid} and
Lemma~\ref{lem:group_ordering}, there also exists an $\eps$-grid bounded representation $\calR'$
having the order in the indistinguishable groups as defined above. So for each component $C_t$, one
of the two orderings $\wlt$ constructed for the linear programs agrees with the left-to-right order
of $C_t$ in $\calR'$.

We want to show that the representation of each component $C_t$ in $\calR'$ gives a solution to one of
the two linear programs associated to $C_t$. We denote by $E'_t$ the value of $E_t$ in the
representation $\calR'$, and by $E^{\min}_t$ the value of $E_t$ obtained by the algorithm after
solving the two linear programming problems associated to $C_t$. We show by induction on $t$
that $E^{\min}_t \le E'_t$, which specifically implies that $E^{\min}_t$ exists and at least one of
the linear programs for $C_t$ is solvable.

We start with $C_1$. As argued above, the left-to-right order in $\calR'$ agrees with one of the
orderings $\wlt$, so the representation of $C_1$ satisfies the constraints~(\ref{constr_ordering}).
Since $E_0=-\infty$, the lower bounds are not modified. By Lemma~\ref{lem:constraints}, the rest of
the constraints are also satisfied. Thus the representation of $C_1$ gives a feasible solution for
the program and gives $E^{\min}_1 \le E'_1$.

Assume now that, for some $C_t$ with $t \ge 1$, at least one of the two linear programming problems
associated to $C_t$ admits a solution, and from induction hypothesis we have $E^{\min}_t \le E'_t$.
In $\calR'$, two neighboring components are represented at distance at least $\eps$. Therefore for
every vertex $v_i$ of $C_{t+1}$, it holds $\ell_i \ge E'_t+\eps \ge E^{\min}_t+\eps$, so the
modification of the lower bound constraints is satisfied by $\calR'$. Similarly as above using
Lemma~\ref{lem:constraints}, the representation of $C_{t+1}$ in $\calR'$ satisfies the remaining
constraints. It gives some solution to one of the programs and we get $E^{\min}_{t+1} \le E'_{t+1}$.

In summary, if there exists a bounded representation, for each component $C_t$ at least one of the
two linear programming problems associated to $C_t$ admits a solution. Therefore, the algorithm
returns a correct bounded representation $\calR$ (as discussed in the beginning of the proof).
We note that $\calR$ does not have to be an $\eps$-grid representation since the linear program just
states that non-intersecting intervals are at distance at least $\eps$. To construct an $\eps$-grid
representation if necessary, we can proceed as in the proof of Lemma~\ref{eps_grid}.\qed
\end{proof}

We note that it is possible to reduce the number of constraints of the linear program from $\O(k^2)$
to $\O(k)$, since neighbors of each $v_i$ appear according to Lemma~\ref{pint_characterization}
consecutively in $\wlt$.  Using the ordering constraints~(\ref{constr_ordering}), we can replace
constraints~(\ref{constr_edges}) and~(\ref{constr_nonedges}) by a linear number of constraints as
follows. For each $v_j$, there are two cases. If $v_j$ is adjacent to all vertices $v_i$ such that
$v_i \wlt v_j$, then we only state the constraint~(\ref{constr_edges}) for $v_{1}$ and $v_j$.
Otherwise, let $v_i$ be the rightmost vertex such that $v_i \wlt v_j$ and $v_iv_j\notin E$. Then we
only state the constraint~(\ref{constr_edges}) for $v_{i+1}$ and $v_j$, and the
constraint~(\ref{constr_nonedges}) for $v_i$ and $v_j$. This is equivalent to the original
formulation of the problem.

In general, any linear program can be solved in $\O(n^{3.5}r^2 \log r \log \log r)$ time by using
Karmarkar's algorithm~\cite{KarmarkarLP}. However, our linear program is special which allows to use
faster techniques:

\begin{proposition}
The \brep\ problem with prescribed $\blt$ can be solved in time $\O(n^2r+nD(r))$.
\end{proposition}

\begin{proof}
Without loss of generality, we assume that the upper and lower bounds restrict the final
representation (if it exists) to lie in the interval $[1,n+3]$. For a given $i$, let $j_i$ be the
index such that $v_{j_i}$ is the rightmost neighbor of $v_i$ in $\wlt$. Let $h_i$ be the index
such that $v_{h_i}$ is the rightmost vertex such that $v_{h_i} \wlt v_i$ and $v_{h_i}v_i\notin E$.
(Notice that $h_i$ might not be defined, in which case we ignore inequalities containing it.)

We replace the variables $\ell_1,\dots,\ell_k$ by $x_0,\dots,x_k$ such that $\ell_i = x_i-x_0$. We
want to solve the following linear system:
\begin{alignat*}{3}
&\text{Minimize:}\qquad&E_t :\!&= x_k-x_0+1,\qquad\ \,&&\nonumber\\
&\text{subject to:}\qquad&x_i-x_{i+1} &\le 0,&\qquad&\forall i = 1,\dots,k-1,\\
&&x_0-x_i &\le -\lbound(v_i),&&\forall i = 1,\dots,k,      \\
&&x_i-x_0 &\le \ubound(v_i),&&\forall i = 1,\dots,k,     \\
&&x_{j_i}-x_i &\le 1,&&  \forall i = 1,\dots,k,        \\
&&x_{h_i}-x_i  &\le  -1-\eps,&& \forall i = 1,\dots,k.
\end{alignat*}
The obtained linear program is a \emph{system of difference constraints}, since each inequality has
the form $x_i-x_j\leq b_{i,j}$.

\begin{figure}[t!]
\centering
\includegraphics{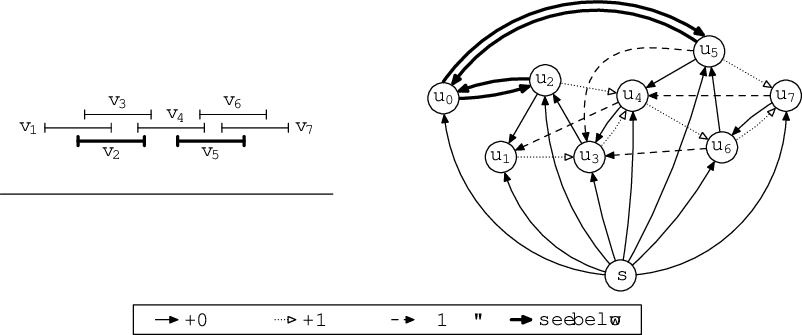}
\caption{On the left, a unit interval graph with two pre-drawn intervals. On the right, the
corresponding digraph $D$ with the weight encoded as in the box. The weights of the bold edges are
as follows:
$w(u_0,u_2) = \ubound(v_2)$, $w(u_0,u_5) = \ubound(v_5)$, $w(u_2,u_0) = -\lbound(v_2)$, and
$w(u_5,u_0) = -\lbound(v_5)$.}
\label{fig:difference_constraints_example}
\end{figure}

Following~\cite[Chapter 24.4]{introalgo}, if the system is feasible, a solution, which is not
necessarily optimal, can be found as follows.  We define a weighted digraph $D$ as follows. As the
vertices, we have $V(D)=\{s,u_0,u_1,\ldots, u_k\}$ where $u_i$ corresponds to $x_i$ and $s$ is a
special vertex. For the edges $\vec E(D)$, we first have an edge $(s,u_i)$ of the weight zero for
every $u_i$. Then for every constraint $x_i-x_j \le b_{i,j}$, we add the edge $(u_j,u_i)$ of the
weight $b_{i,j}$. See Figure~\ref{fig:difference_constraints_example}.

As proved in~\cite[Chapter 24.4]{introalgo}, there are two possible cases.  If $G$ \emph{contains} a
negative-weight cycle, then there is no feasible solution for the system.  If $G$ \emph{does not
contain} negative-weight cycles, then we define $\delta(s,u_i)$ as the weight of the minimum-weight
path connecting $s$ to $u_i$ in $G$.  Then we put $x_i = \delta(s,u_i)$ for each $i$ which defines a
feasible solution of the system.  Moreover, this solution minimizes the objective function
$\max\{x_i\}-\min \{x_i\}$. We next show that this function is equivalent to the objective function
in our linear program.

Suppose that we have a solution of our system, satisfying the constraints but not necessarily
optimizing the objective function.  Because of our assumption that the representation lies in the
interval $[1,n+3]$, we know that $\ell_i>0$ for all $i$. Therefore, $x_i>x_0$. So $\min \{x_i\}$ is
always attained by $x_0$, while $\max\{x_i\}$ is always attained by $x_k$. So minimization of the
objective function $\max\{x_i\}-\min \{x_i\}$ is equivalent to the original minimization of $E_t =
x_k-x_0+1$.

In order to find a negative-weight cycle in $D$ or, alternatively, compute the weight of the
minimum-weight paths from $s$ to all the other vertices of $D$, we use the Bellman-Ford algorithm.
Notice that Dijkstra's algorithm cannot be used in this case, since some edges of $D$ have negative
weight. We next analyze the running time of the whole procedure.

We assume that the cost of arithmetic operations with large numbers is not constant. The algorithm
computes the value $\eps$ in the beginning which can be clearly done in time $\O(nD(r))$.  (Instead
of the least common multiple we can simply compute the product of $q_i$'s.)

Afterwards, we compute the weights of the edges of $D$ as multiples of $\eps$, which takes time
$\O(kD(r))$. Then each step of the Bellman-Ford algorithm requires time $\O(r)$, and the algorithm
runs $\O(k^2)$ steps in total. The total time to solve each linear program is therefore
$\O(k^2r+kD(r))$. Finally, the total time of the algorithm is $\O(n^2r+nD(r))$.\qed
\end{proof}

In the next subsections, we improve the time complexity of the \brep\ problem with prescribed $\blt$
to $\O(n^2 + nD(r))$. Our algorithm makes use of several structural properties of the set of all
representations. We note that structural properties of the polyhedron of our linear program, in the
case where all lower bounds equal zero and there are no upper bounds, have been considered in
several papers in the context of semiorders~\cite{semiorder_minimal_rep,semiorder_polytopes}.

\subsection{The Partially Ordered Set $\rset$} \label{subsec:poset_rset}

Let the graph $G$ in consideration be a connected unit interval graph. We study structural
properties of its representations. Suppose that we fix one of the two partial left-to-right orders
$<$ of the intervals from Section~\ref{sec:proper_interval_graphs}, so that only indistinguishable
vertices are incomparable. We also fix some positive $\eps = {1 \over K}$. For most of this section,
we work just with lower bounds and completely ignore upper bounds.

We define $\rset$ as the set of all $\eps$-grid representations satisfying the lower bounds and in
some left-to-right ordering that extends $<$. We define a very natural partial ordering $\le$ on
$\rset$: We say that $\calR \le \calR'$ if and only if $\ell_i \le \ell'_i$ for every $v_i \in
V(G)$; i.e., $\le$ is the carthesian ordering of vectors $(\ell_1,\dots,\ell_n)$. In this section,
we study structural properties of the poset $(\rset,\le)$.

If $\eps \le {1 \over n}$, then $\rset \ne \emptyset$. The reason is that the graph $G$ is a unit
interval graph, and thus there always exists an $\eps$-grid representation $\calR$ far to the right
satisfying the lower bound contraints.

\heading{The Semilattice Structure.}
Let us assume that $\lbound(v_i) > -\infty$ for some $v_i \in V(G)$.  Let $S$ be a subset of
$\rset$. The infimum $\inf(S)$ is the greatest representation $\calR \in \rset$ such that $\calR \le
\calR'$ for every $\calR' \in S$. In a general poset, infimums may not exist, but if they exist,
they are always unique. For $\rset$, we show:

\begin{lemma} \label{lem:semilattice}
Every non-empty $S \subseteq \rset$ has an infimum $\inf(S)$.
\end{lemma}

\begin{proof}
We construct the requested infimum $\cal R$ as follows:
$$\ell_i = \min \{ \ell'_i : \calR' \in S\},\qquad \forall v_i \in V(G).$$
Notice that the positions in $\calR$ are well-defined, since the position of each interval in each
$\cal R'$ is bounded and always on the $\eps$-grid. Clearly, if $\calR$ is a correct representation,
it is the infimum $\inf(S)$. It remains to show that $\calR \in \rset$.

Clearly, all positions in $\calR$ belong to the $\eps$-grid and satisfy the lower bound constraints.
Let $v_i$ and $v_j$ be two vertices. The values $\ell_i$ and $\ell_j$ in $\calR$ are given by two
representations $\calR_1,\calR_2 \in S$, that is, $\ell_i = \ell^1_i$ and $\ell_j = \ell^2_j$.
Notice that the left-to-right order in $\calR$ has to extend $<$: If $v_i < v_j$, then $\ell_i =
\ell^1_i \le \ell^2_i < \ell_j^2 = \ell_j$, since $\calR_1$ minimizes the position of $v_i$ and the
left-to-right order in $\calR_2$ extends $<$.  Concerning correctness of the representation of the
pair $v_i$ and $v_j$, we suppose that $\ell_i = \ell^1_i \le \ell^2_j = \ell_j$; otherwise we swap
$v_i$ and $v_j$.
\begin{packed_itemize}
\item First we suppose that $v_iv_j \in E(G)$. Then $\ell^2_j \le \ell^1_j$, since $\calR_2$
minimizes the position of $v_j$. Since $\calR_1$ is a correct representation, $\ell^1_j - 1 \le
\ell^1_i$. So $\ell_j - 1 \le \ell_i \le \ell_j$, and the intervals $v_1$ and $v_2$ intersect.
\item The other case is when $v_iv_j \notin E(G)$. Then $\ell^1_i \le \ell^2_i \le \ell^2_j - 1 -
\eps$, since $\calR_1$ minimizes the position of $v_i$, $\calR_2$ is a correct representation and
$v_i < v_j$ in both representations.  So $v_i$ and $v_j$ do not intersect in $\calR$ as requested.
\end{packed_itemize}
Consequently, $\calR$ represents correctly each pair $v_i$ and $v_j$, and hence $\calR \in
\rset$.\qed
\end{proof}

A poset is a \emph{(meet)-semilattice} if every pair of elements $a,b$ has an infimum
$\inf(\{a,b\})$. Lemma~\ref{lem:semilattice} shows that the poset $(\rset,\le)$ forms a
(meet)-semilattice.  Similarly as $\rset$, we could consider the poset set of all ($\eps$-grid)
representations satisfying both the lower and the upper bounds. The structure of this poset is a
\emph{complete lattice}, since all subsets have infimums and supremums. Lattices and semilattices
are frequently studied, and posets that are lattices satisfy very strong algebraic properties.

\heading{The Left-most Representation.}
We are interested in a specific representation in $\rset$, called the \emph{left-most
representation}. An $\eps$-grid representation $\calR \in \rset$ is the left-most representation if
$\calR \le \calR'$ for every $\calR' \in \rset$; so the left-most representation is left-most in
each interval at the same time. We note that the notion of the left-most representation does not
make sense if we consider general representations (not on the $\eps$-grid). The left-most
representation is the infimum $\inf(\rset)$, and thus by Lemma~\ref{lem:semilattice} we get:

\begin{corollary} \label{cor:lftmost_rep}
The left-most representation always exists and it is unique.
\end{corollary}

There are two algorithmic motivations for studying left-most representations. First, in the linear
program of Section~\ref{subsec:lp_approach} we need to find a representation minimizing $E_t$.
Clearly, the left-most representation is minimizing $E_t$ and in addition it is minimizing the rest
of the endpoints as well. The second motivation is that we want to construct a representation
satisfying the upper bounds as well, so it seems reasonable to try to place every interval as far to
the left as possible. The left-most representation is indeed a good candidate for a bounded
representation:

\begin{lemma} \label{lem:just_lbounds}
There exists a representation $\calR'$ satisfying both lower and upper bound constraints if and only if the
left-most representation $\calR$ satisfies the upper bound constraints.
\end{lemma}

\begin{proof}
Since $\calR \in \rset$, it satisfies the lower bounds. If $\calR$ satisfies the upper bound
constraints, it is a bounded representation. On the other hand, let $\calR'$ be a bounded
representation. Then
$$\lbound(v_i) \le \ell_i \le \ell'_i \le \ubound(v_i),\qquad \forall v_i \in V(G),$$
and the left-most representation is also a bounded representation.\qed
\end{proof}

\subsection{Why Left-most Representations Cannot Be Easily Constructed by Iterations?}

A very natural idea for an algorithm is to construct the left-most representation iteratively, by
adding the vertices $v_1,\dots,v_n$ one by one and recomputing the left-most representation in each
step.  In this section, we describe why this natural algorithm does not run in quadratic time. More
precisely, we do not claim that it is not possible to implement it in quadratic time or faster using
some additional tricks and structural results, but we did not succeeded in this matter.

\heading{The Iterative Algorithm.}
Let $G$ be a connected unit interval graph, and let $<$ be the left-to-right partial ordering of its
vertices $v_1,\dots,v_n$ numbered from left to right. We denote by $G_k$ the graph induced by
$\{v_1,\dots,v_k\}$. Let $\calR_k$ be the left-most representation of $G_k$, and let $\ell_i^k$ be
the position of the left endpoint of $v_i$ in $\calR_k$. The iterative algorithm runs as follows.

We initiate $\calR_1$ with $\ell^1_1 = \lbound(v_1)$. To compute $\calR_k$ from $\calR_{k-1}$, we
first put $\ell^k_i := \ell^{k-1}_i$ for all $1 \le i \le k-1$, and $\ell^k_k :=
\max\bigl\{\lbound(v_k),\ell^k_j+1+\eps\bigr\}$ where $v_j$ is the rightmost placed non-neighbor of
$v_k$. Since $\calR_k$ is not likely a correct representation of $G_k$, we proceed by a series of
\emph{fixes} till we obtain a correct representation:
\begin{packed_itemize}
\item If $v_iv_j \in E(G_k)$, $i<j$, and $\ell^k_i < \ell^k_j - 1$, we \emph{fix} $\calR_k$ by
setting $\ell^k_i := \ell^k_j - 1$.
\item If $v_iv_j \notin E(G_k)$, $i<j$, and $\ell^k_i \ge \ell^k_j - 1$, we \emph{fix} $\calR_k$ by
setting $\ell^k_j := \ell^k_i + 1 + \eps$.
\end{packed_itemize}

\heading{Correctness.} We start by proving that the above algorithm is correct.

\begin{proposition} \label{prop:iterative_correctness}
The above iterative algorithm stops after finite number of steps and outputs the left-most
representation $\calR$.
\end{proposition}

\begin{proof}
It is just sufficient to show that it constructs the left-most representation $\calR_k$ from the
left-most representation $\calR_{k-1}$, and the rest is true by induction. Let $\calR^s_k$ be a
vector of positions created by the algorithm after $s$ fixes, so $\calR^s_k$ might not be a correct
representation. We prove by induction according to $s$ that $\calR^s_k \le \calR_k$.

Since $\calR_{k-1}$ is the left-most representation of $G_{k-1}$, we get $\calR_{k-1} \le \calR_k
|_{G_{k-1}}$. We initiate $\ell^k_k$ as far to the left as possible, and thus $\calR^0_k \le \calR_k$.
Now let $\calR^{s-1}_k \le \calR_k$. Then we easily get $\calR^s_k \le \calR_k$ since the fix of
$(v_i,v_j)$ shifts one of them as little to the right as necessary; since $\calR_k$ is a correct
representation, it clearly cannot have the shifted interval more to the left than $\calR^s_k$.

Since each fix strictly increases the position of one interval and according to
Corrolary~\ref{cor:lftmost_rep} the left-most representation $\calR_k$ always exists, we cannot
apply fixes indefinitely and the algorithm outputs some correct representation $\calR^s_k$. Since
$\calR^s_k \le \calR_k$, we get $\calR^s_k = \calR_k$.\qed
\end{proof}

\heading{Unclear Complexity.} Even though the above algorithm is correct, it is not even clear that
its complexity is polynomial in $n$ and does not depend on $\eps$. We did not try to further
estimate this complexity but it seems one could bound the number of fixes in each iteration by
something like $\O(n^2)$ which would give a cubic-time algorithm. The reason why this does not give
a quadratic-time algorithm is that the position of each interval can be updated by multiple fixes.
We always shift as little as possible, and not as much as it is required by the structure of the graph.
Furthermore, we simplified our analysis by assuming that we can locate a wrongly represented pair $(v_i,v_j)$
in constant time, and that we compute on the arithmetic machine (so we ignored numerical issues with
small values of $\eps$).

Nevertheless, we believe that the complexity of this algorithm could be improved which might lead to
a different quadratic-time (or potentially even linear-time) algorithm for the bounded
representation problem with prescribed ordering $\wlt$. As a good starting point, we suggest that
one should get a good structural understand how much $\calR_k$ differs from $\calR_{k-1}$. Even
through we give some additional properties concerning the left-most representation, we still do not
fully understand its structure. Therefore we derived a different algorithm based on shifting which
we describe in the rest of Section~\ref{sec:prescribed_brep}.

\subsection{Left-Shifting of Intervals} \label{subsec:left_shifting}

\begin{figure}[b]
\centering
\includegraphics{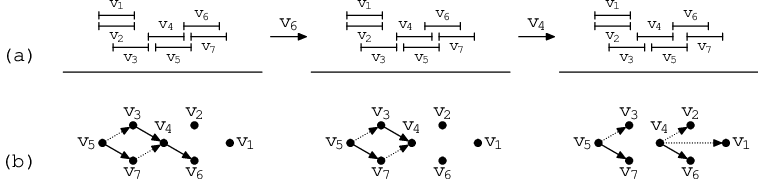}
\caption{(a) A representation modified by left-shifting of $v_6$ and $v_4$.
(b) The corresponding obstruction digraphs $H$ for each of the representations. Only sinks of the
obstruction digraphs can be left-shifted.}
\label{fig:digraph_H}
\end{figure}

Suppose that we construct some initial $\eps$-grid representation that is not the left-most
representation. We want to transform this initial representation in $\rset$ into the left-most
representation of $\rset$ by applying a sequence of the following simple operations called the
\emph{left-shifting}.  The left-shifting operation shifts one interval of the representations by
$\eps$ to the left such that this shift maintains the correctness of the representation; for an
example see~Figure~\ref{fig:digraph_H}a. The main goal of this section is to prove that by
left-shifting we can always produce the left-most representation.

\begin{proposition} \label{prop:shifting_prop}
For $\eps = {1 \over K}$ and $K \ge {n \over 2}$, an $\eps$-grid representation $\calR \in \rset$ is
the left-most representation if and only if it is not possible to shift any single interval to the
left by $\eps$ while maintaining correctness of the representation.
\end{proposition}

Before proving the proposition, we describe some additional combinatorial structure of
left-shifting. An interval $v_i$ is called \emph{fixed} if it is in the left-most
position and cannot ever be shifted more to the left, i.e., $\ell_i = \min\{\ell'_i : \calR' \in
\rset\}$. For example, an interval $v_i$ is fixed if $\ell_i = \lbound(v_i)$. A representation is
the left-most representation if and only if every interval is fixed.

\heading{Obstruction Digraph.}
An interval $v_i$, having $\ell_i > \lbound(v_i)$, can be left-shifted if it does
not make the representation incorrect, and the incorrectness can be obtained in two ways.
First, there could be some interval $v_j$, $v_j \wlt v_i$ such that $v_iv_j \notin
E(G)$ and $\ell_j+1+\eps = \ell_i$; we call $v_j$ a \emph{left obstruction} of $v_i$. Second, there
could be some interval $v_j$, $v_i \wlt v_j$ such that $v_iv_j \in E(G)$ and $\ell_i+1 = \ell_j$ (so
$v_i$ and $v_j$ are touching); then we call $v_j$ a \emph{right obstruction} of $v_i$. In both
cases, we first need to move $v_j$ before moving $v_i$.

For the current representation $\calR$, we define the \emph{obstruction digraph} $H$ on the
vertices of $G$ as follows. We put $V(H) = V(G)$ and $(v_i,v_j) \in E(H)$ if and only if $v_j$ is an
obstruction of $v_i$. For an edge $(v_i,v_j)$, if $v_j \wlt v_i$, we call it a \emph{left edge}; if
$v_i \wlt v_j$, we call it a \emph{right edge}. As we apply left-shifting, the structure of $H$
changes; see Figure~\ref{fig:digraph_H}b.

\begin{lemma} \label{lem:fixed_intervals}
An interval $v_i$ is fixed if and only if there exists a directed path in $H$ from $v_i$ to $v_j$
such that $\ell_j = \lbound(v_j)$.
\end{lemma}

\begin{proof}
Suppose that $v_i$ is connected to $v_j$ by a path in $H$.  By the definition of $H$, $v_xv_y \in E(H)$
implies that $v_y$ has to be shifted before $v_x$. Thus $v_j$ has to be shifted before moving $v_i$ which
is not possible since $\ell_j = \lbound(v_j)$.

On the other hand, suppose that $v_i$ is fixed, i.e., $\ell_i = \inf\{\ell'_i: \forall {\cal R'}\}$.
Let $H'$ be the induced subgraph of $H$ of the vertices $v_j$ such that there exists a directed path from
$v_i$ to $v_j$. If for all $v_j \in H'$, $\ell_j > \lbound(v_j)$, we can shift all vertices of $H'$
by $\eps$ to the left which constructs a correct representation and contradicts that $v_i$ is fixed.
Therefore, there exists $v_j \in H'$ having $\ell_j = \lbound(v_j)$ as requested.\qed
\end{proof}

For example in Figure~\ref{fig:digraph_H} on the left, if $\ell_4 = \lbound(v_4)$, then the
intervals $v_3$, $v_4$, $v_5$ and $v_7$ are fixed. Also, we can prove:

\begin{lemma} \label{lem:digraph_acyclic}
If $\eps = {1 \over K}$ and $K \ge {n \over 2}$, the obstruction digraph $H$ is acyclic.
\end{lemma}

\begin{proof}
Suppose for contradiction that $H$ contains some cycle $u_1,\dots,u_c$. This cycle contains $a$ left
edges and $b$ right edges. Recall that if $(u_i,u_{i+1})$ is a left edge, then $\ell_{u_{i+1}} =
\ell_{u_i} - 1 - \eps$, and if it is a right edge, $\ell_{u_{i+1}} = \ell_{u_i} + 1$ (and similarly
for $(u_c,u_1)$). If we go along the cycle from $u_1$ to $u_1$, the initial and the final positions
have to be the same. Therefore $a(1+\eps) = b$.

Now if this equation holds, then $a$ has to be a multiple of $K$. Therefore $a \ge K$ and $b \ge
K+1$, and thus $n \ge c = a+b \ge 2K+1$ which is not possible.\qed
\end{proof}

\begin{figure}[t]
\centering
\includegraphics{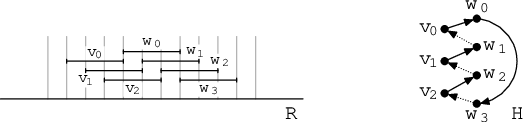}
\caption{An $\eps$-grid representation for $\eps = {1 \over 3}$ on the left and the obstruction
digraph $H$ containing a cycle on the right.}
\label{fig:acyclic_counterexample}
\end{figure}

We note that the assumption $K \ge {n \over 2}$ is necessary and tight. For every $\eps = {1 \over K}$, there
exists a representation of a graph with $2K+1$ vertices having a cycle in $H$. The graph contains
two cliques $v_0,\dots,v_{K-1}$ and $w_0,\dots,w_K$ such that $v_i$ is also adjacent to
$w_0,\dots,w_i$. Then the assignment $\ell_{v_0} = 0$, $\ell_{v_i} = \ell_{v_0} + i\eps$ and
$\ell_{w_i} = \ell_{v_0} + 1 + i\eps$ is a correct representation. Observe that $H$ contains a
cycle $w_kv_{k-1}w_{k-1}v_{k-2}w_{k-2}\dots v_1w_1v_0w_0w_k$. See Figure~\ref{fig:acyclic_counterexample} for $K = 3$.

\heading{Predecessors of Poset $\rset$.}
A representation $\calR' \in \rset$  is a \emph{predecessor} of $\calR \in \rset$ if $\calR' <
\calR$ and there is no representation $\bar\calR \in \rset$ such that $\calR' < \bar\calR < \calR$. We
denote the predecessor relation by $\prec$. In a general poset, predecessors may not
exist. But they always exist for a poset of a discrete structure like $(\rset,\le)$: Indeed, there
are finitely many representations $\bar\calR$ between any $\calR' < \calR$, and thus the
predecessors always exist. Also, for any two
representations $\calR' < \calR$, there exists a finite \emph{chain} of predecessors $\calR' = \calR_0 \prec
\calR_1 \prec \cdots \prec \calR_k = \calR$.

For the poset $(\rset,\le)$, we are able to fully describe the predecessor structure:

\begin{lemma} \label{lem:prec_relation}
For $\eps = {1 \over K}$ and $K \ge {n \over 2}$, the representation $\calR'$ is a predecessor of
$\calR$ if and only if $\calR'$ is obtained from $\calR$ by applying one left-shifting operation.
\end{lemma}

\begin{proof}
Clearly, if $\calR'$ is obtained from $\calR$ by one left-shifting, it is a predecessor of $\calR$.

On the other hand, suppose we have $\calR' < \calR$. Let $H$ be the obstruction digraph of $\calR$
and $\bar H$ be the subgraph of $H$ induced by the intervals having different positions in $\calR$
and $\calR'$.  Then there are no directed edges from $\bar H$ to $H \setminus \bar H$ (otherwise
$\calR'$ would be an incorrect representation). According to Lemma~\ref{lem:digraph_acyclic}, the
digraph $\bar H$ is acyclic. Therefore, it contains at least one sink $v_i$. By left-shifting $v_i$
in $\calR$, we create a correct representation $\bar\calR \in \rset$. Clearly, $\calR' \le \bar\calR
\prec \calR$, and so $\calR'$ is a predecessor of $\calR$ if and only if $\calR' = \bar\calR$.\qed
\end{proof}

Again, the assumption on the value of $\eps$ is necessary. For example in
Figure~\ref{fig:acyclic_counterexample}, the structure of $\rset$ is just a single chain where a
predecessor of some representation is obtained by shifting all intervals by $\eps$ to the left.

\heading{Proof of Left-shifting Proposition.} The main proposition of this subsection is a simple
corollary of Lemma~\ref{lem:prec_relation}.

\begin{proof}[Proposition~\ref{prop:shifting_prop}]
The left-most representation $\calR$ is $\inf(\rset)$, so it has no predecessors and nothing can be
left-shifted. On the other hand, if $\inf(\rset) < \calR$, there is a chain of predecessors in between which
implies using Lemma~\ref{lem:prec_relation} that it is possible to left-shift some interval.\qed
\end{proof}

\subsection{Preliminaries for the Shifting Algorithm} \label{subsec:prunning}

Before describing the shifting algorithm, we present several results which simplify the graph
and the description of the algorithm.

\heading{Pruned Graph.}
The obstruction digraph $H$ may contain many edges since each vertex $v_i$ can have many
obstructions. But if $v_i$ has many, say, left obstructions, these obstructions have to be
positioned the same. If two intervals $u$ and $v$ have the same position in a correct unit interval
representation, then $N[u] = N[v]$ and they are indistinguishable. Our goal is to construct a
\emph{pruned graph} $G'$ which replaces each group of indistinguishable vertices of $G$ by a single
vertex. This construction is not completely straightforward since indistinguishable vertices may
have different lower and upper bounds.

Let $\{\Gamma_1,\dots,\Gamma_k\}$ be the partitioning of $V(G)$ by the groups of indistinguishable vertices
(and the groups are ordered by $\wlt$ from left to right). We construct a unit interval graph $G'$,
where the vertices are $\gamma_1,\dots,\gamma_k$ with $\lbound(\gamma_i) = \max\{\lbound(v_j) : v_j
\in \Gamma_i\}$, and the edges $E(G')$ correspond to the edges between the groups of $G$.

\begin{figure}[b]
\centering
\includegraphics{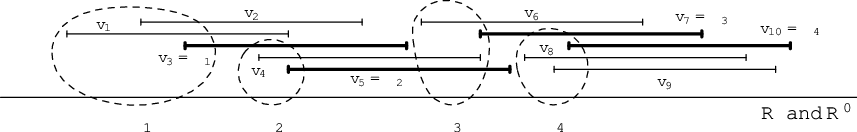}
\caption{Both representations $\calR$ and $\calR'$ in one figure, with the intervals of $\calR'$
depicted in bold. The left endpoints of the intervals of each group are enclosed by dashed curves, and
these curves are ordered from left to right according to $<$.}
\label{fig:both_representations}
\end{figure}

Suppose that we have the left-most representation $\calR'$ of the pruned graph $G'$ and we want to
construct the left-most representation $\calR$ of $G$. Let $\Gamma_\ell$ be a group. We place each
interval $v_i \in \Gamma_\ell$ as follows.  Let $\gamma^\ell_\leftarrow$ be the first non-neighbor of
$\gamma_\ell$ on the left and $\gamma^\ell_\rightarrow$ be the right-most neighbor of $\gamma_\ell$ (possibly
$\gamma^\ell_\rightarrow = \gamma_\ell$). We set
\begin{equation} \label{placing_position}
\ell_i = \max\{\lbound(v_i),\ell_{\gamma^\ell_\leftarrow}+1+\eps,\ell_{\gamma^\ell_\rightarrow}-1\},
\end{equation}
and if $\gamma^\ell_\leftarrow$ does not exist, we ignore it in $\max$. The meaning of this formula is to place
each interval as far to the left as possible while maintaining the structure of $\calR'$.
Figure~\ref{fig:both_representations} contains an example of the construction of $\calR$.

Before proving correctness of the construction of $\calR$, we show two general properties of the
formula~(\ref{placing_position}). The first lemma states that each interval $v_i \in \Gamma_\ell$ is
not placed in $\calR$ too far from the position of $\gamma_\ell$ is $\calR'$.

\begin{lemma} \label{lem:important_inequal}
For each $v_i \in \Gamma_\ell$, it holds
\begin{equation} \label{important_inequal}
\ell_{\gamma_\ell} - 1 \le \ell_i \le \ell_{\gamma_\ell}.
\end{equation}
\end{lemma}

\begin{proof}
The first inequality is true since $\ell_{\gamma_\ell} - 1 \le \ell_{\gamma^\ell_\rightarrow} -1 \le
\ell_i$ holds according to (\ref{placing_position}) and the ordering $\wlt$ for $\calR'$. The second
inequality holds since $\calR'$ is a correct bounded representation, and so $\ell_{\gamma_\ell}$ is
greater than or equal to each term in~(\ref{placing_position}).\qed
\end{proof}

The second lemma states that the representations $\calR$ and $\calR'$ are \emph{intertwining} each
other. If $\calR$ is drawn on top of $\calR'$, then the vertices of each group $\Gamma_\ell$ are in
between of $\gamma_{\ell-1}$ and $\gamma_\ell$; see Figure~\ref{fig:both_representations}.

\begin{lemma} \label{lem:intertwining}
For each $v_i \in \Gamma_\ell$ and $\ell > 1$, it holds
\begin{equation} \label{intertwining}
\ell_{\gamma_{\ell-1}} < \ell_i \le \ell_{\gamma_\ell},
\end{equation}
\end{lemma}

\begin{proof}
The second inequality holds by~(\ref{important_inequal}). For the first inequality,
there are two possible cases why the groups $\Gamma_{\ell-1}$ and $\Gamma_\ell$ are distinct:
\begin{packed_itemize}
\item The first case is when $\gamma^{\ell}_\leftarrow$ is a neighbor of $\gamma_{\ell-1}$.
Then $\ell_{\gamma_{\ell-1}} \le \ell_{\gamma_\leftarrow^\ell} + 1 < \ell_i$; the first inequality
holds since $\gamma_\leftarrow^\ell\gamma_{\ell-1} \in E(G')$ and $\calR'$ is a correct
representation, and the second inequality is given by~(\ref{placing_position}).
\item The second case is when $\gamma^{\ell}_\rightarrow$ is a non-neighbor of $\gamma_{\ell-1}$.
Then $\ell_{\gamma_{\ell-1}} < \ell_{\gamma_\rightarrow^\ell}-1 \le \ell_i$ by the fact that
$\gamma_{\ell-1}\gamma_\rightarrow^\ell \notin E(G')$ and by~(\ref{placing_position}).
\end{packed_itemize}
In both cases, we get $\ell_{\gamma_{\ell-1}} < \ell_i$.\qed
\end{proof}

Now, we are ready to show correctness of the construction of $\calR$.

\begin{proposition} \label{prop:pruned_graph}
From the left-most representation $\calR'$ of the pruned graph $G'$, we can construct the correct
left-most representation $\calR$ of $G$ by placing the intervals according to (\ref{placing_position}).
\end{proposition}

\begin{proof}
We argue the correctness of the representation $\calR$. Let $v_i$ and $v_j$ be a pair of vertices of
$G$. Let $v_iv_j \in E(G)$. If $v_i$ and $v_j$ belong to the same group $\Gamma_\ell$, they intersect
each other at position $\ell_{\gamma_\ell}$ by~(\ref{important_inequal}).
Otherwise let $v_i \in \Gamma_\ell$ and $v_j \in \Gamma_{\ell'}$, and assume that $\Gamma_\ell <
\Gamma_{\ell'}$. Then $\ell_i \le \ell_{\gamma_\ell} \le \ell_j$ by the intertwining
property~(\ref{intertwining}). Also, $\ell_j \le \ell_{\gamma_{\ell'}} \le \ell_{\gamma^{\ell}_\rightarrow} \le \ell_i+1$
since $\gamma_{\ell'}$ is a right neighbor of $\gamma_\ell$ and~(\ref{important_inequal}).
Therefore, $\ell_i \le \ell_j \le \ell_i+1$ and $v_i$ intersects $v_j$ in $\calR$.
Now, let $v_iv_j \notin E(G)$, $v_i \in \Gamma_\ell$, $v_j \in \Gamma_{\ell'}$ and $v_i <
v_j$.  Then $\ell_i \le \ell_{\gamma_\ell} \le \ell_{\gamma^{\ell'}_\leftarrow} \le \ell_j - 1 -
\eps$ by~(\ref{placing_position}) and~(\ref{important_inequal}), so $v_i$ and $v_j$ do not
intersect. So the assignment $\calR$ is a correct representation of $G$.

It remains to show that $\calR$ is the left-most representation of $G$.  We can identify each
$\gamma_\ell$ with one interval $v_i \in \Gamma_\ell$ having $\lbound(v_i) = \lbound(\gamma_\ell)$;
for an example see Figure~\ref{fig:both_representations}. So $G'$ can be viewed as an induced
subgraph of $G$. We want to show that the intervals of $G'$ are represented in $\calR$ exactly the
same as in $\calR'$. Since $\calR |_{G'}$ (which denotes $\calR$ restricted to $G'$) is some
representation of $G'$ and $\calR'$ is the left-most representation of $G'$, we get $\ell'_{\gamma_\ell}
\le \ell_{\gamma_\ell}$ for every $\gamma_\ell$. By~(\ref{important_inequal}), we get
$\ell'_{\gamma_\ell} = \ell_{\gamma_\ell}$.

We know that $\calR |_{G'}$ is the left-most representation, or in other words each interval of $G'$
is fixed in $\calR$. The rest of the intervals are placed so that they are either trivially fixed
by $\ell_i = \lbound(v_i)$, or they have as obstructions some fixed intervals from $G'$, in which
case they are fixed by Lemma~\ref{lem:fixed_intervals}. Therefore, every interval of $G$ is fixed and
$\calR$ is the left-most representation.\qed
\end{proof}

For the pruned graph $G'$, the obstruction digraph $H$ has in- and out-degree at most two. Each
interval has at most one left obstruction and at most one right obstruction, and these obstructions
are always the same intervals. More precisely, if $v_j$ is a left obstruction of $v_i$, then $v_j =
v_\leftarrow^i$, whereas if $v_j$ is a right obstruction of $v_i$, then $v_j = v_\rightarrow^i$.

The pruning operation can be done in time $\O(n+m)$, so we may assume that our graph $G$ is already
pruned and contains no indistinguishable vertices. And the structure of obstructions in $G$ can be
computed in time $\O(n+m)$ as well.

\heading{Position Cycle.}
For each interval in some $\eps$-grid representation, we can write its position in this form:
\begin{equation} \label{pos_def}
\ell_i = \alpha_i + \beta_i\eps,\qquad \alpha_i \in {\mathbb Z},\ \beta_i \in {\mathbb Z}_K,
\end{equation}
where $\eps = {1 \over K}$. In other words, $\alpha_i$ is the integer position of $v_i$ in the grid
and $\beta_i$ describes how far is this interval from this integer position.

\begin{figure}[t]
\centering
\includegraphics{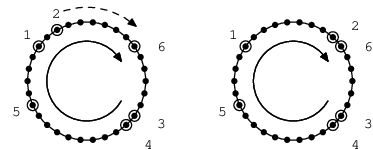}
\caption{Examples of position cycles. In the cycle on the left, we can shift $\beta_2$ in the clockwise
direction towards $\beta_6$, which gives a new representation whose position cycle is depicted on
the right. We note that after this left-shifting, $v_6$ is not necessarily an obstruction of $v_2$.}
\label{fig:position_cycle}
\end{figure}

Concerning left-shifting, the values $\beta_i$ are more important. We can depict ${\mathbb Z}_K =
\{0,\dots,K-1\}$ as a cycle with $K$ vertices where the value decreases clockwise. The value
$\beta_i$ assigns to each interval $v_i$ one vertex of the cycle. The cycle ${\mathbb Z}_K$ together
with marked positions of $\beta_i$'s is called the \emph{position cycle}.  A vertex of the position
cycle is called \emph{taken} if some $\beta_i$ is assigned to it, and \emph{empty} otherwise.  The
position cycle allows us to visualize and work with left-shifting very intuitively.  When an
interval $v_i$ is left-shifted, $\beta_i$ cyclically decreases by one, so $\beta_i$ moves clockwise
along the cycle.  For an illustration, see Figure~\ref{fig:position_cycle}.

If $(v_i,v_j)$ is a left edge of $H$, then $\beta_j = \beta_i-1$, and if $(v_i,v_j)$ is a right edge,
then $\beta_i = \beta_j$.  So if $v_j$ is an obstruction of $v_i$, $\beta_j$ has to be very close to
$\beta_i$ (either at the same position or at the next clockwise position). If there is a big empty
space in the clockwise direction from $\beta_i$, the interval $v_i$ can be left-shifted many times
(or till it becomes fixed by $\ell_i = \lbound(v_i)$). Notice that if $\beta_i$ is very close to
$\beta_j$, it does not mean that $\ell_i$ is very close to $\ell_j$ because the values $\alpha_i$ and
$\alpha_j$ are ignored in the position cycle.

\subsection{The Shifting Algorithm for \brep}
\label{subsec:quadratic_algo}

We want to solve an instance of \brep\ with a prescribed ordering $\blt$. We work with an
$\eps$-grid which is different from the one in Section~\ref{subsec:eps_grid}. The new value of
$\eps$ is the value given by~(\ref{eps_def}) refined $n$ times, so
$$\eps = {1 \over n^2} \cdot \eps'.$$
Lemma~\ref{eps_grid} applies for this value of $\eps$ as well, so if the instance is solvable, there
exists a solution which is on this $\eps$-grid.

The algorithm works exactly as the algorithm of Subsection~\ref{subsec:lp_approach}. The only
difference is that for a component with $k$ vertices we can solve the linear program in time
$\O(k^2+kD(r))$, and now we describe how to do it. We assume that the input component is already
pruned, otherwise we prune it and use Proposition~\ref{prop:pruned_graph} to complete the
representation. We expect that the left-to-right order $\wlt$ of the vertices is given.  The
algorithm requires time $\O(kD(r))$ since the bounds are given in the form $p_i \over q_i$ and we need
to perform arithmetic operations with these bounds.  Therefore the total complexity of the algorithm
for the \brep\ problem is $\O(n^2+nD(r))$.

\heading{Overview.} The algorithm for solving one linear program works in three basic steps:
\begin{packed_enum}
\item We construct an initial $\eps$-grid representation (in the ordering $\wlt$) having $\ell_i \ge
\lbound(v_i)$ for all intervals, using the algorithm of Corneil et al.~\cite{uint_corneil}.
\item We shift the intervals to the left while maintaining correctness of the representation until
the left-most representation is constructed, using Proposition~\ref{prop:shifting_prop}.
\item We check whether the left-most representation satisfies the upper bounds.
If so, we have the left-most representation satisfying all bound constraints. This representation
solves the linear program of Subsection~\ref{subsec:lp_approach} and minimizes $E_t$. Otherwise, the
left-most representation does not satisfy the upper bound constraints. Thus by
Lemma~\ref{lem:just_lbounds} no representation satisfies the upper bound constraints, and
the linear program has no solution.
\end{packed_enum}

\heading{Input Size.} Let $r$ be the size of the input describing bound constraints.
A standard complexity assumption is that we can operate with polynomially large numbers
(having $\O(\log r)$ bits in binary) in constant time, to avoid the extra factor $\O(\log r)$ in
the complexity of most of the algorithms. However, the value of $\eps$ given by~(\ref{eps_def})
might require $\O(r)$ digits when written in binary. The assumption that we can computate with
numbers having $\O(r)$ digits in contant time would break most of the computational models.
Therefore, our computational model requires a larger time for arithmetic operations with numbers
having $\O(r)$ digits in binary. For example, the best known algorithm for multiplication/division
on a Turing machine requires time $\O(D(r))$.

The problem is that a straightforward implementation of our algorithm working with the $\eps$-grid
would require time $\O(k^2r^c)$ for some $c$ instead of $\O(k^2+kD(r))$. There is an easy way out.
Instead of computing with \emph{long numbers} having $\O(r)$ digits, we mostly compute with
\emph{short numbers} having just $\O(\log r)$ digits. Instead of the $\eps$-grid, we mostly work in
a larger $\Delta$-grid where $\Delta = {1 \over n^2}$. The algorithm computes with
the long numbers only in two places. First, some initial computations concerning the input are
performed. Second, when the shifting makes some interval fixed, the algorithm estimes the final
$\eps$-grid position of the interval. All these computations can be done in total time $\O(kD(r))$
and we describe everything in detail later.

\heading{Left-Shifting.}
The basic operation of the algorithm is the \leftshift\ procedure which we describe here.  We deal
separately with fixed and unfixed intervals (and some intervals might be fixed initially).  Unfixed
intervals are on the $\Delta$-grid and fixed intervals have precise positions calculated on the
$\eps$-grid. We place only unfixed intervals on the position cycle for the $\Delta$-grid.  At any
moment of the algorithm, each vertex of the position cycle is taken by at most one $\beta_i$; this
is true for the initial representation and the shifting keeps this property.

We define the procedure $\leftshift(v_i)$ which shifts $v_i$ from the position $\ell_i$ into a new
position $\ell'_i$ such that the representation remains correct. The procedure $\leftshift(v_i)$
consists of two steps:
\begin{packed_enum}
\item Since $v_i$ is unfixed, it has some $\beta_i$ placed on the position cycle. Let $k$ be such
that the vertices $\beta_i+1,\dots,\beta_i+k$ of the position cycle are empty and the vertex
$\beta_i+k+1$ is taken by some $\beta_b$. Then a candidate for the new position of $v_i$ is
$\bar\ell_i = \ell_i-k\Delta$.
\item We need to ensure that this shift from $\ell_i$ to $\bar\ell_i$ is valid with respect to
$\lbound(v_i)$ and the positions of the fixed intervals.  Concerning the lower bound, we cannot
shift further than $\lbound(v_i)$. Concerning the fixed intervals, the shift is limited by positions
of fixed obstructions of $v_i$. If $v_j$ is a fixed left obstruction, we cannot shift further than
$\ell_j+1+\eps$, and if $v_{j'}$ a fixed right obstruction, we cannot shift further than
$\ell_{j'}-1$.
\end{packed_enum}
The resulting position after applying $\leftshift(v_i)$ is
\begin{equation} \label{new_position}
\ell'_i = \max\{\bar\ell_i, \lbound(v_i), \ell_j+1+\eps, \ell_{j'}-1\}.
\end{equation}

\begin{lemma} \label{lem:lshift_procedure}
If the original representation $\calR$ is correct, than the $\leftshift(v_i)$ procedure produces a
correct representation $\calR'$.
\end{lemma}

\begin{proof}
Clearly, the lower bound for $v_i$ is satisfied in $\calR'$. The shift of $v_i$ from $\ell_i$ to
$\ell'_i$ can be viewed as a repeated application of the left-shifting operation from
Section~\ref{subsec:left_shifting}. We just need to argue that each left-shifting operation can be
applied till the position $\ell'_i$ is reached.

If at some point, the left-shifting operation could not be applied, there would have to be some
obstruction $v_j$ of $v_i$. There is no unfixed obstruction since all vertices of the position cycle
$\beta_i+1,\dots,\beta_i+k$ are empty. And $v_j$ cannot be fixed as well since we check positions of
both possible obstructions. So there is no obstruction $v_j$. Therefore, by repeated applying the
left-shifting operation, the interval $v_i$ gets at a position $\ell'_i$ and the resulting
representation is correct.\qed
\end{proof}

After $\leftshift(v_i)$, if $\bar\ell_i$ is not a strict maximum of the four terms
in~(\ref{new_position}), the interval $v_i$ becomes fixed; either trivially since $\ell'_i =
\lbound(v_i)$, or by Lemma~\ref{lem:fixed_intervals} since $v_i$ becomes obstructed by some fixed
interval.  In such a case, we remove $\beta_i$ from the position cycle.

\heading{Fast Implementation of Left-Shifting.}
Since we apply the \leftshift\ procedure repeatedly, we want to implement it in time $\O(1)$.
Considering the terms in~(\ref{new_position}), the first term $\bar\ell_i$ is a short number (on
the $\Delta$-grid) and the remaining terms are long numbers (on the $\eps$-grid). We first compare
$\bar\ell_i$ to the remaining terms which are three comparisons of short and long numbers and we are
going to show how to compare them in $\O(1)$. If $\bar\ell_i$ is a strict maximum, we use it for
$\ell'_i$. Otherwise, we need to compute the maximum of the remaining three terms which takes time
$\O(D(r))$. But then the interval $v_i$ becomes fixed, and so this costly step is done exactly $k$
times, and takes the total time $\O(kD(r))$.

\begin{lemma} \label{lem:fast_shifting}
With the total precomputation time $\O(kD(r))$, it is possible to compare $\bar\ell_i$ to the
remaining terms in~(\ref{new_position}) in time $\O(1)$ per \leftshift\ procedure.
\end{lemma}

\begin{proof}
Initially, we do the following precomputation for the lower bounds. By the input, we have $b$ lower
bounds given in the form ${p_1 \over q_1},\dots,{p_b \over q_b}$ as irreducible fractions. For each
bound, we first compute its position $(\alpha_i,\beta_i)$ on the $\eps$-grid; see~(\ref{pos_def}).

If $\lbound(v_i) \ll \lbound(v_j)$ for some vertices $v_i$ and $v_j$, then $\lbound(v_i)$ is never
achieved since the graph is connected and every representation takes space at most $k$. Therefore we
can increase $\lbound(v_i)$ without any change in the solution of the instance. More precisely, let
$\alpha = \max \alpha_i$.  Then we modify each bound by setting $\alpha_i :=
\max\{\alpha-k-1,\alpha_i\}$. In addition, we shift all the bounds by substructing a constant $C$ such
that each $\alpha_i - C \in [0,k+1]$.  Concerning $\beta_i$, we round the position
$(\alpha_i,\beta_i)$ down to a position $(\alpha_i,\bar\beta_i)$ of the $\Delta$-grid. These
precomputations can be done for all lower bounds in time $\O(kD(r))$.

Suppose that we want to find out whether $\bar\ell_i \le \lbound(v_j) = \alpha_j + \beta_j\cdot
\eps$ where $\bar\ell_i$ is in the $\Delta$-grid. Then it is sufficient to check whether $\bar\ell_i
\le \alpha_j+\bar\beta_j\Delta$ which can be done in constant time since both $\alpha_j$ and
$\bar\beta_j$ are short numbers.

When $v_j$ becomes fixed, its precise position is computed using~(\ref{new_position}). Then we
compute the values $\ell_j-1$ and $\ell_j+1+\eps$ used in~(\ref{new_position}) and round them down
to the $\Delta$-grid. Using these precomputed values, $\bar\ell_i$ can be compared with the
remaining terms in~(\ref{new_position}) in time $\O(1)$.  When an interval becomes fixed, time
$\O(D(r))$ is used.  Since each interval becomes fixed exactly once, this rounding also takes the total
time $\O(kD(r))$.\qed
\end{proof}

Notice that the representation is constructed in a position shifted by $C$. Later, before checking
the upper bound, we shift the whole representation back.

\heading{Initial Representation.}
Recall that the position cycle has $n^2$ vertices and $\Delta = {1 \over n^2}$.  The algorithm of
Corneil et al.~\cite{uint_corneil} gives a representation in the $1 \over k$-grid. Using the proof
of Lemma~\ref{eps_grid}, we construct from it the initial $\Delta$-grid representation. Then we
shift it such that $\ell_i \ge \lbound(v_i)$ for each $v_i$ and $\ell_i \le \lbound(v_i) + \Delta$
for some $v_i$. For this initial representation, each interval can be shifted to the left in total
by at most $\O(k)$.

\begin{figure}[b]
\centering
\includegraphics{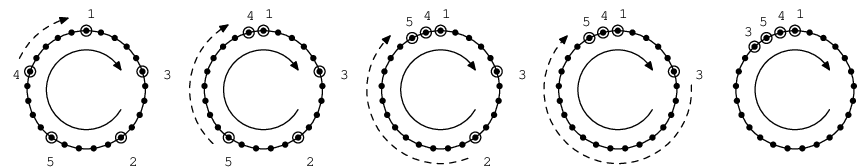}
\caption{The position cycle during the first phase, changing from left to right. The first phase
clusters the $\beta_i$'s by moving $\beta_4$, $\beta_5$, $\beta_2$ and $\beta_3$ towards $\beta_1$.
When $\leftshift(v_2)$ is applied, $v_2$ becomes fixed and $\beta_2$ disappears from the position
cycle.}
\label{fig:first_phase}
\end{figure}

The initial representation obtained from the representation of the algorithm of Corneil et
al.~\cite{uint_corneil} places all intervals in such a way that $\beta_i$'s are almost positioned
equidistantly in the position cycle; refer to the left-most position cycle in
Figure~\ref{fig:first_phase}. As we say in the description of the \leftshift\ procedure, we only
require that all $\beta_i$'s are placed to pairwise different vertices of the position cycle.

\begin{figure}[t]
\centering
\includegraphics{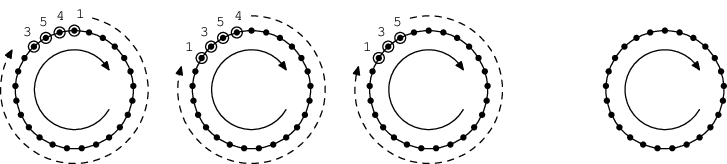}
\caption{The position cycle during the second phase, changing from left to right. We shift $\beta_i$'s
across the big gap till all $\beta_i$'s disappear.}
\label{fig:second_phase}
\end{figure}

\heading{Shifting Phases.}
All shifting of the algorithm is done by repeated application of the \leftshift\ procedure. Using
Lemma~\ref{lem:lshift_procedure}, we know that the representation created in each step is correct.
We apply the procedure in such a way that each interval is almost always shifted by almost one.
The shifting of unfixed intervals proceeds in two phases:
\begin{packed_itemize}
\item \emph{The first phase} creates one big gap by clustering all $\beta_i$'s in one part of the cycle. To
do so, we apply the \leftshift\ procedure to each interval, in the order given by the position
cycle. Of course, some intervals might become fixed and disappear from the position cycle. We
obtain one big gap of size at least $n(n-1)$. Again, refer to Figure~\ref{fig:first_phase}.
\item \emph{In the second phase,} we use this big gap to shift intervals one by one, which also moves the
cluster along the position cycle. Again, if some interval becomes fixed, it is removed from the
position cycle. The second phase finishes when each interval becomes fixed and the left-most
representation is constructed. For an example, see Figure~\ref{fig:second_phase}.
\end{packed_itemize}

\heading{Putting It All Together.}
First, we show correctness of the shifting algorithm and its complexity:

\begin{lemma} \label{lem:shifting_algorithm}
For a component having $k$ vertices, the shifting algorithm constructs a correct left-most
representation in time $\O(k^2+kD(r))$.
\end{lemma}

\begin{proof}
First, we argue correctness of the algorithm. The algorithm starts with an initial representation
which is correct and satisfies the lower bounds. By Lemma~\ref{lem:lshift_procedure}, after applying each
\leftshift\ procedure, the resulting representation is still correct. The algorithm keeps a correct
list of fixed intervals which is increased by shifting. So after finitely many applications of the
\leftshift\ procedure, every interval becomes fixed, and we obtain the left-most representation.

Concerning complexity, all precomputations take total time $\O(kD(r))$. Using
Lemma~\ref{lem:fast_shifting}, each $\leftshift(v_i)$ procedure can be applied in time $\O(1)$
unless $v_i$ becomes fixed.  The first phase is applying the \leftshift\ procedure $k-1$ times. In
the second phase, each interval is shifted by at least $n-1 \over n$ (unless it becomes fixed).
Since each interval can be shifted by at most $\O(k)$ from its initial position, the second phase
applies the \leftshift\ procedure $\O(k^2)$ times. So the total running time of the algorithm is
$\O(k^2+kD(r))$.\qed
\end{proof}

We are ready to prove that \brep\ with a prescribe ordering $\blt$ can be solved in time
$\O(n^2+nD(r))$:

\begin{proof}[Theorem~\ref{thm:quadratic_brep}]
We proceed exactly as in the algorithm of Section~\ref{subsec:lp_approach}, so we process the
components $C_1 \blt \cdots \blt C_c$ from left to right, and for each of them we solve two linear
programs. For each linear program, we find the left-most representation using
Lemma~\ref{lem:shifting_algorithm}, and we test for this representation (shifted back by $C$)
whether the upper bounds are satisfied. According to Lemma~\ref{lem:just_lbounds}, the linear
program is solvable if and only if the left-most representation satisfy the upper bounds, and
clearly the left-most representation minimizes $E_t$. The time complexity of the algorithm is
$\O(n^2+nD(r))$ and the proof of correctness is exactly the same as in
Proposition~\ref{prop:shifting_prop}.\qed
\end{proof}

We finally present an FPT algorithm for $\brep$ with respect to the number of components $c$. The
algorithm is based on Theorem~\ref{thm:quadratic_brep}.

\begin{proof}[Corollary~\ref{cor:brep_fpt}]
There are $c!$ possible left-to-right orderings of the components of $G$. For each of them,
we can decide in time $\O(n^2+nD(r))$ whether there exists a bounded representation in the order, using
Theorem~\ref{thm:quadratic_brep}. So the total time necessary is $\O((n^2+nD(r))c!)$.\qed
\end{proof}


\section{Extending Unit Interval Graphs} \label{sec:unit_interval_graphs}

The $\ext(\uint)$ problem can be solved using Theorem~\ref{thm:quadratic_brep}.
We just need to show that it is a particular instance of \brep\ in which the ordering $\blt$ of the
components can be derived:

\begin{proof}[Theorem~\ref{cor:unit}]
The graph $G$ contains unlocated components and located components. Similarly to
Section~\ref{sec:proper_interval_graphs}, unlocated components can be placed far to the right and we
can deal with them using a standard recognition algorithm.

Concerning located components $C_1,\dots,C_c$, they have to be ordered in $\calR'$ from left to
right, which gives the required ordering $\blt$. We straightforwardly construct the instance of
\brep\ with this $\blt$ as follows.  For each pre-drawn interval $v_i$ at position $\ell_i$, we put
$\lbound(v_i) = \ubound(v_i) = \ell_i$. For the rest of the intervals, we set no bounds. Clearly,
this instance of \brep\ is equivalent with the original $\ext(\uint)$ problem. And we can solve it
in time $\O(n^2+nD(r))$ using Theorem~\ref{thm:quadratic_brep}.\qed
\end{proof}


\section{Conclusions}\label{sec:concl}

\heading{Assumption on the Input.}
Almost every graph algorithm is not able to achieve time $\O(n+m)$ if the input is given by an
adjacency matrix of the graph.  Similarly, to get linear time in Theorem~\ref{thm:pint}, we have to
assume that the partial representation of a proper interval graph is given in a nice form.

We say that a partial representation is \emph{normalized} if the pre-drawn endpoints have
positions $\{1,\ldots,2n\}$. This assumption is natural since according to Lemma~\ref{lem:pint_extendible},
the extendibility of a partial representation only depends on the left-to-right order of the
pre-drawn intervals and not on the precise positions.  For a normalized partial representation, the
order $<^{G'}$ can be computed in time $\O(n)$.  If the representation is not given in this way, the
algorithm needs an additional time $\O(k \log k)$ to construct $<^{G'}$, where $k$ is the number of
pre-drawn intervals.

\heading{Polyhedron Interpretation.}
Consider the linear program of Section~\ref{subsec:lp_approach}.  The described shifting algorithm
has the following geometric interpretation. When the constraints~(\ref{constr_ub}) are omitted, all
solutions of the linear program form an unbounded polyhedron. The initial solution is one point of
the polyhedron and the left-most representation is the vertex of the polyhedron minimizing all
values $\ell_i$. One application of the \leftshift\ procedure corresponds to decreasing one variable
while staying in the polyhedron.  The algorithm computes a Manhatten-like path from the initial
solution to the left-most representation consisting of $\O(n^2)$ shifts.

We believe that the polyhedron has some additional useful structure which might be exploited for
constructing faster algorithms and might lead to discovering new useful properties of unit interval
representations. It is also an interesting question whether some of our techniques can be
generalized to other systems of difference constraints.

\heading{Simultaneous Representations.} Let $G_1,\dots,G_k$ be graphs having
$V(G_i) \cap V(G_j) = I$ for each $i \ne j$. The $\simrep(\calC)$ problem asks whether there exists
representations $\calR_1,\dots,\calR_k$ of $G_1,\dots,G_k$ (of class $\calC$) which assign the same
sets to the vertices of $I$. This problem was considered in~\cite{simultaneous_interval_graphs} and
its relations to the partial representation extension problem were discussed in~\cite{kkv,blas_rutter}.

We believe that it is possible to apply results and techniques to solve these problems for proper
and unit interval graphs. First, one needs to construct simultaneous left-to-right orderings
$<_1,\dots,<_k$ having the same order on $I$. Then, we can use linear programming/shifting approach
to construct the simultaneous representation. This is a possible direction of future research.

\heading{Open Problem.} To conclude the paper, we present two open problems.
\begin{problem}
Is it possible to solve the problem $\ext(\uint)$ in faster time than $\O(n^2 + nD(r))$?
\end{problem}

We consider the other problem as currently the major open problem concerning restricted
representations of graphs. The class of the intersection graphs of arcs of a circle is called
\emph{circular-arc graphs} (\ca); for references see~\cite{egr}. We ask the following question:

\begin{problem}
Can the problem $\ext(\ca)$ be solved in polynomial time?
\end{problem}

We believe that solving this problem might lead to a better understanding of the class itself. All
known polynomial-time recognition algorithms are quite complex, and construct specific types of
representations called \emph{canonical representations}. Further, many results concerning
circular-arc graphs were later shown to be false; for instance recently the graph isomorphism
problem of circular-arc graphs is again open. To solve $\ext(\ca)$, the structure of all
representations needs to be better understood which could be a major breakthrough concerning this
and other classes.

\bibliographystyle{elsarticle-num}
\bibliography{extending_pint_and_uint}

\begin{thebibliography}{10}
\expandafter\ifx\csname url\endcsname\relax
  \def\url#1{\texttt{#1}}\fi
\expandafter\ifx\csname urlprefix\endcsname\relax\def\urlprefix{URL }\fi
\expandafter\ifx\csname href\endcsname\relax
  \def\href#1#2{#2} \def\path#1{#1}\fi

\bibitem{kkorssv}
P.~Klav\'{\i}k, J.~Kratochv\'{\i}l, Y.~Otachi, I.~Rutter, T.~Saitoh,
  M.~Saumell, T.~Vysko\v{c}il, Extending partial representations of proper and
  unit interval graphs, in: Algorithm Theory -- SWAT 2014, Vol. 8503 of Lecture
  Notes in Computer Science, 2014, pp. 253--264.

\bibitem{hajos_interval_graphs}
G.~Haj{\'o}s, {\"U}ber eine {A}rt von {G}raphen, Internationale Mathematische
  Nachrichten 11 (1957) 65.

\bibitem{gilmore64}
P.~C. Gilmore, A.~J. Hoffman, A characterization of comparability graphs and of
  interval graphs, Can. J. Math. 16 (1964) 539--548.

\bibitem{PQ_trees}
K.~S. Booth, G.~S. Lueker, Testing for the consecutive ones property, interval
  graphs, and planarity using {PQ}-tree algorithms, J. Comput. System Sci. 13
  (1976) 335--379.

\bibitem{LBFS_int}
D.~G. Corneil, S.~Olariu, L.~Stewart, The {LBFS} structure and recognition of
  interval graphs, SIAM J. Discrete Math. 23~(4) (2009) 1905--1953.

\bibitem{agt}
M.~C. Golumbic, Algorithmic Graph Theory and Perfect Graphs, North-Holland
  Publishing Co., 2004.

\bibitem{egr}
J.~P. Spinrad, Efficient Graph Representations, Field Institute Monographs,
  2003.

\bibitem{kkv}
P.~Klav\'{\i}k, J.~Kratochv\'{\i}l, T.~Vysko\v{c}il, Extending partial
  representations of interval graphs, in: Theory and Applications of Models of
  Computation, TAMC 2011, Vol. 6648 of Lecture Notes in Computer Science, 2011,
  pp. 276--285.

\bibitem{blas_rutter}
T.~Bl{\"a}sius, I.~Rutter, Simultaneous {PQ}-ordering with applications to
  constrained embedding problems, in: SODA'13: Proceedings of the Twenty-Fourth
  Annual ACM-SIAM Symposium on Discrete Algorithms, 2013, pp. 1030--1043.

\bibitem{kkosv}
P.~Klav\'{\i}k, J.~Kratochv\'{\i}l, Y.~Otachi, T.~Saitoh, T.~Vysko\v{c}il,
  Linear-time algorithm for partial representation extension of interval
  graphs, In preparation.

\bibitem{kkkw}
P.~Klav{\'i}k, J.~Kratochv{\'i}l, T.~Krawczyk, B.~Walczak, Extending partial
  representations of function graphs and permutation graphs, in: Algorithms,
  ESA 2012, Vol. 7501 of Lecture Notes in Computer Science, 2012, pp. 671--682.

\bibitem{cfk}
S.~Chaplick, R.~Fulek, P.~Klavík, Extending partial representations of circle
  graphs, in: Graph Drawing, Vol. 8242 of LNCS, Springer, 2013, pp. 131--142.

\bibitem{kkos}
P.~Klav\'{\i}k, J.~Kratochv\'{\i}l, Y.~Otachi, T.~Saitoh, Extending partial
  representations of subclasses of chordal graphs, in: Algorithms and
  Computation, ISAAC 2012, Vol. 7676 of Lecture Notes in Computer Science,
  2012, pp. 444--454.

\bibitem{int_planar_hard}
S.~Chaplick, P.~Dorbec, J.~Kratochv\'{\i}l, M.~Montassier, J.~Stacho, Contact
  representations of planar graph: Rebuilding is hard, To appear in WG 2014.

\bibitem{simultaneous_interval_graphs}
K.~R. Jampani, A.~Lubiw, Simultaneous interval graphs, in: Algorithms and
  Computation, ISAAC 2010, Vol. 6506 of Lecture Notes in Computer Science,
  2010, pp. 206--217.

\bibitem{jl-srpcc-12}
K.~R. Jampani, A.~Lubiw, The simultaneous representation problem for chordal,
  comparability and permutation graphs, Journal of Graph Algorithms and
  Applications 16~(2) (2012) 283--315.

\bibitem{proper_is_unit}
F.~S. Roberts, Indifference graphs, in: F. Harary (Ed.), Proof Techniques in
  Graph Theory, Academic Press, 1969, pp. 139--146.

\bibitem{angelini}
P.~Angelini, G.~D. Battista, F.~Frati, V.~Jel\'{\i}nek, J.~Kratochv\'{\i}l,
  M.~Patrignani, I.~Rutter, Testing planarity of partially embedded graphs, in:
  SODA'10: Proceedings of the Twenty-First Annual ACM-SIAM Symposium on
  Discrete Algorithms, 2010, pp. 202--221.

\bibitem{patrignani}
M.~Patrignani, On extending a partial straight-line drawing, Int. J. Found.
  Comput. Sci. 17~(5) (2006) 1061--1070.

\bibitem{bko}
M.~Balko, P.~Klavík, Y.~Otachi, Bounded representations of interval and proper
  interval graphs, in: Algorithms and Computation, Vol. 8283 of LNCS, Springer,
  2013, pp. 535--546.

\bibitem{introalgo}
T.~H. Cormen, C.~E. Leiserson, R.~L. Rivest, C.~Stein, Introduction to
  Algorithms, Third Edition, 3rd Edition, The MIT Press, 2009.

\bibitem{division}
M.~F\"{u}rer, Faster integer multiplication, SIAM J. Comput. 39~(3) (2009)
  979--1005.

\bibitem{deng}
X.~Deng, P.~Hell, J.~Huang, Linear-time representation algorithms for proper
  circular-arc graphs and proper interval graphs, SIAM J. Comput. 25~(2) (1996)
  390--403.

\bibitem{recog_chordal_graphs}
D.~J. Rose, R.~E. Tarjan, G.~S. Lueker, Algorithmic aspects of vertex
  elimination on graphs, SIAM Journal on Computing 5~(2) (1976) 266--283.

\bibitem{roberts_phd_thesis}
F.~S. Roberts, Representations of indifference relations, Ph.D. Thesis,
  Stanford University, 1968.

\bibitem{uint_corneil}
D.~G. Corneil, H.~Kim, S.~Natarajan, S.~Olariu, A.~P. Sprague, Simple linear
  time recognition of unit interval graphs, Inform. Process. Lett. 55~(2)
  (1995) 99--104.

\bibitem{partition}
M.~R. Garey, D.~S. Johnson, Complexity results for multiprocessor scheduling
  under resource constraints, SIAM J. Comput. 4~(4) (1975) 397--411.

\bibitem{KarmarkarLP}
N.~Karmarkar, A new polynomial-time algorithm for linear programming,
  Combinatorica 4~(4) (1984) 373--395.

\bibitem{semiorder_minimal_rep}
M.~Pirlot, Minimal representation of a semiorder, Theory and Decision 28 (1990)
  109--141.

\bibitem{semiorder_polytopes}
B.~Balof, J.~P. Doignon, S.~Fiorini, The representation polyhedron of a
  semiorder, Order 30~(1) (2013) 103--135.

\end{thebibliography}

\end{document}